\documentclass[%
preprint,
 amsmath,amssymb,
aps,
prb,
]{revtex4-1}

\usepackage{color}
\usepackage{boxedminipage}
\usepackage{enumerate}
\usepackage{subcaption}
\usepackage{caption}
\usepackage{epsf}
\usepackage{indentfirst}
\usepackage{esvect}
\usepackage{bbold}
\usepackage{dsfont}
\usepackage{braket}
\usepackage{graphicx}
\usepackage{dcolumn}
\usepackage{bm}

\newcommand{\beq}{\begin{equation}}
\newcommand{\eeq}{\end{equation}}
\newcommand{\nnbeq}{\begin{equation*}}
\newcommand{\nneeq}{\end{equation*}}
\newcommand{\bo}{\textbf}
\newcommand{\lv}{\lvert}
\newcommand{\rv}{\rvert}

\begin{document}

\title{A Hartree-Fock study of the \(\nu=0\) quantum Hall state of monolayer graphene with short range interactions}

\author{Braden Feshami}
 \altaffiliation[]{}
\author{H. A. Fertig}%
 \email{bfeshami@indiana.edu}
\affiliation{%
 Indiana University\\
Physics Department
}%

\date{\today}

\begin{abstract}
 Recent experiments involving tilted graphene samples have shown evidence of a continuous phase transition in the $\nu=0$ quantum Hall bulk state. We present here a simple model that supports such a transition. In addition to a long range SU(4) symmetric Coulomb interaction, we include Hubbard on-site and nearest neighbor interactions with tunable coupling strengths, and perform a self-consistent Hartree-Fock analysis. A large sea of negative energy Landau levels is retained, and is shown to have important qualitative and quantitative effects. Phase diagrams are constructed within the space of physically relevant parameters, yielding results consistent with experimental observation.
\begin{description}
\item[PACS numbers]

\end{description}
\end{abstract}

\pacs{Valid PACS appear here}
\maketitle

\section{Introduction}
Graphene, a two-dimensional (2D) honeycomb lattice of carbon atoms, has proven itself to be a remarkable material, and serves as an excellent platform to investigate electrons confined to two dimensions. Shortly after the material was isolated \cite{Novoselov2004}, clear evidence of the integer quantum Hall sequence was observed \cite{Kim2005,Novoselov2005}. The orbital degrees of freedom for non-interacting electrons in graphene are governed by a massless Dirac equation, and, in the presence of a magnetic field, this gives rise to a relativistic Landau Level (LL) spectrum, $\epsilon(n)=sgn(n)\frac{\hbar v_F}{\ell}\sqrt{2\lv n \rv}$. Here, $v_F$ is the Fermi velocity ($\sim 10^8 cm/s)$, $\ell=\sqrt{\frac{\hbar}{eB_{\perp}}}$ is the magnetic length, $n$ is a positive or negative integer, and $B_{\perp}$ is the perpendicular component of the magnetic field. The presence of negative energy LL's is a consequence of the relativistic nature of the effective Hamiltonian, which distinguishes the spectrum from that of conventional 2D electron systems, as does the square-root dependence on $B_{\perp}$. In addition to orbital degrees of freedom, electrons in graphene host discrete degrees of freedom, the spin and valley indices, yielding four distinct LLs for each value of $n$.

One of the unique behaviors of graphene is its support of a $\nu=0$ quantum Hall effect, which occurs when as many LL's are occupied as are unoccupied, in which case the system is nominally undoped. In the non-interacting limit, this entails occupying two of the four $n=0$ levels, the choice of which being determined by the Zeeman coupling. Early transport experiments \cite{Kim2007} on monolayer graphene at $\nu=0$ were consistent with this. However, subsequent measurements \cite{Young2012} on higher quality samples were not, strongly suggesting that interactions cannot be ignored in this system. 

The nature of the $\nu=0$ groundstate has thus become the topic of a number of studies. The simplest models, in which interactions are treated as SU(4) symmetric in the spin and valley indices \cite{DasSarma2006}, cannot explain this behavior, since only the Zeeman coupling is left to break the symmetry, yielding a ferromagnetic state essentially the same as the non-interacting one. Because of the underlying lattice structure, however, interactions need not be truly SU(4) symmetric \cite{Basko2008,Kharitonov2012}, and the inclusion of these effects leads to other possible phases.

A seminal study \cite{Kharitonov2012} of this system, which includes only the $n=0$ LL's within a Hamiltonian with two phenomenological parameters that describe the possible symmetry-breaking of the interactions (SU(4) $\rightarrow$ SU(2)$\otimes$U(1)), demonstrated that, in addition to the ferromagnetic (FM) state, the system can host a canted antiferromagnetic (CAF) state, a charge density (CDW) ordered state, and a Kekul\'{e} dimerized (KD) state. While all of these are insulating in the bulk, the FM state is distinguished at its edge by the presence of a helical edge mode \cite{Levitov2006,Fertig2006,Fertig2009,Fertig2014,Fertig2016a,Fertig2016b}. This allows for transport in systems with edges, while the other phases are expected to be insulating.

An explicit investigation of this possibility was reported in Ref. \onlinecite{Young2014}, which discussed the results of tilted field experiments, allowing the field perpendicular to the sample ($B_{\perp}$) to be smaller than the total field ($B_T$). Because the orbital degrees of freedom are sensitive to the former, whereas the Zeeman coupling is proportional to the latter, the Zeeman energy may be greatly enhanced in such experiments. The experiment demonstrated a change from insulating to conducting behavior with increasing $B_T$ and fixed $B_{\perp}$, which can most naturally be understood as a transition from the CAF to the FM state \cite{Kharitonov2012,Young2014,Fertig2014,Fertig2016a,Fertig2016b}. However, whether this transition can happen for realistic interaction parameters remains a subject of debate\cite{Herbut2007,Miransky2006,Miransky2008,DasSarma2014,FernandoRossier2014,Shankar2016}.

In what follows, we seek to better understand the nature of this insulating groundstate, and explore the phase diagram using a model with microscopically meaningful interactions, specific to graphene. In addition to an SU(4) symmetric (long-range) Coulomb interaction, Hubbard-like on-site (OS) and nearest neighbor (NN) interactions are included as lattice scale corrections. The strengths of the short range interactions are tuned by the parameters $V_0$ (OS) and $V_1$ (NN). Our approach is a numerical Hartree-Fock (HF) approximation, in which negative energy LL's, and LL-mixing (LLM) terms, are explicitly included. We find that the negative energy sea plays an essential role in determining which phases appear in the phase diagram, and that, within our model, the  (KD) phase does not appear. The model does support a continuous phase transition between the CAF and FM states, as described above. and, moreover, we find this to provide a quantitatively reasonable explanation for the experimental observations. Lastly, our results are in agreement with a recent variational study \cite{Shankar2016}.

This article is organized as follows. In Section 2, the model Hamiltonian is presented, and matrix elements for the (HF) direct and exchange components of each interaction are worked out. Contributions from the Dirac sea are also described here. The relevant states, and their density matrix representations, are discussed in Section 3. Numerical results are presented in Section 4, along with phase diagrams, which are constructed for a range of microscopic interaction strengths and Zeeman field strengths.
\section{The Model}
\subsection{Hartree-Fock Hamiltonian}
 Our model Hamiltonian is composed of a single particle (non-interacting) term and an interaction term, $\mathcal{H}=\mathcal{H}_o+\mathcal{H}_{int}$. We use a continuum form for the single particle Hamiltonian, written in terms of the LL spectrum and the Zeeman energy,  
\beq
\mathcal{H}_o=\sum_{n,X,\sigma,\tau}\bigg[sgn(n)\frac{\hbar v_F}{\ell}\sqrt{2\lv n \rv}-\frac{\sigma g^{\ast}\mu_B}{2}B_T\bigg]c^{\dagger}_{nX\sigma\tau}c_{nX\sigma\tau} \label{2.1.1},
\eeq
where $\mu_B$ is the Bohr magneton, $\sigma=+1 \ (-1)$, for spin up (down), $\tau \in \{K,K^{\prime}\}$ labels the two valleys, $g^*=2$, and $X$ is the guiding center quantum number of the electron wavefunction. The total field can be expressed as $B_T=B_{\perp}/cos\theta$, or, in terms of the Zeeman field strength, $\xi_z\equiv1/cos\theta \geq 1$, $B_T=\xi_zB_{\perp}$. The angle $\theta$ measures the tilt of the sample relative to the total magnetic field.

For interactions, we include three contributions, so that $\mathcal{H}_{int}=\mathcal{H}^{(C)}+\mathcal{H}^{(OS)}+\mathcal{H}^{(NN)}$, representing the long range Coulomb, OS, and NN interactions, respectively. The Coulomb interaction preserves the SU(4) symmetry of the Hamiltonian, and does not distinguish between which sublattice, $s\in\{A,B\}$, an electron resides on. The short range interactions, by contrast, have non-trivial sublattice dependence. A generic interaction term takes the form
\beq
\mathcal{H}^{(i)}=\frac{1}{2}\sum_{s,s'}\sum_{\sigma,\sigma'}\int d^2r d^2r' \Psi^{\dagger}_{\sigma}(\bo{r},s)\Psi^{\dagger}_{\sigma'}(\bo{r}',s')\widetilde{V}^{(i)}(\bo{r}-\bo{r}')\Psi_{\sigma'}(\bo{r}',s')\Psi_{\sigma}(\bo{r},s), \label{2.1.2}
\eeq
which we re-express in terms of valley-specific field operators via $\Psi_{\sigma}(\bo{r},s)\equiv\frac{1}{\sqrt{2}}\big[\Psi_{K,\sigma}(\bo{r},s)e^{i\bo{K} \cdot \bo{r}}+\Psi_{K^{\prime},\sigma}(\bo{r},s)e^{i\bo{K}' \cdot \bo{r}} \big]$, with $\bo{K}=-\bo{K}'=\frac{4\pi}{3\sqrt{3}a}\hat{e}_x$, and $a=0.142$nm, the distance between neighboring carbon atoms for unstrained graphene. Substituting this into Eq. \ref{2.1.2}, and dropping any terms which rapidly oscillate, we arrive at
\beq
\mathcal{H}^{(i)}=\frac{1}{2}\sum_{s,s'\sigma,\sigma'}\sum_{\bm{\tau}}\int d^2rd^2r' \Psi^{\dagger}_{\sigma\tau_1}(\textbf{r},s)\Psi^{\dagger}_{\sigma'\tau_2}(\textbf{r}',s')V^{(i)}(\bm{\tau};s,s';\ \textbf{r}-\textbf{r}')\Psi_{\sigma'\tau_3}(\textbf{r}',s')\Psi_{\sigma\tau_4}(\textbf{r},s) \label{2.1.3}.
\eeq
As a matter of notational conveniencce, the bold face on a discrete index will refer to the 4 component set of constituent labels (i.e., $\bm{\tau}=\{\tau_1,\tau_2,\tau_3,\tau_4\}$). Each interaction is specified by its corresponding potential,
\beq
\begin{aligned}
&V^{(C)}(\bo{r}-\bo{r}',\bm{\tau};s,s')=\widetilde{V}^{(C)}(\bo{r}-\bo{r}') \ \delta_{\tau_1,\tau_4}\delta_{\tau_2\tau_3}=\frac{e^2}{4\pi\epsilon_o}\frac{1}{\lv \bo{r}-\bo{r}' \rv}\delta_{\tau_1,\tau_4}\delta_{\tau_2,\tau_3},\\
&V^{(OS)}(\bo{r}-\bo{r}',\bm{\tau};s,s')=\widetilde{V}^{(OS)}(\bo{r}-\bo{r}')f(\bm{\tau})\delta_{s,s'}=\frac{V_0a_c}{4}\delta(\bo{r}-\bo{r}')f(\bm{\tau})\delta_{s,s'},\\
&V^{(NN)}(\bo{r}-\bo{r}',\bm{\tau};s,s')=\widetilde{V}^{(NN)}(\bo{r}-\bo{r}')f(\bm{\tau})\big[1-\delta_{s,s'}\big]=\frac{3V_1a_c}{8}\delta(\bo{r}-\bo{r}')f(\bm{\tau})\big[1-\delta_{s,s'}\big] \label{2.1.4},
\end{aligned}
\eeq
where $a_c=\frac{3\sqrt{3}a^2}{2}$ is the area of a unit cell. $V_0$ and $V_1$ are the OS and NN coupling strengths specific to graphene. Although these values are likely to both be positive, with estimates \cite{Katsnelson2011} putting their magnitudes at a few eVs, their precise values are uncertain, so we treat them as tunable parameters in what follows. The function
\beq
f(\bm{\tau})\equiv \big[\mathbb{1}\big]_{\tau_1\tau_4}\big[\mathbb{1}\big]_{\tau_2\tau_3}+\frac{1}{2}\bigg(\big[\tau_x\big]_{\tau_1\tau_4}\big[\tau_x\big]_{\tau_2\tau_3}+\big[\tau_y\big]_{\tau_1\tau_4}\big[\tau_y\big]_{\tau_2\tau_3}\bigg), \label{2.1.5}
\eeq
where $\tau_x,\tau_y,\tau_z$ are Pauli matrices in the valley space, is non-zero for a total of six possible combinations of valley indices. 

The field operators are further expanded in terms of LL states, $\phi_{\lvert n \rvert,X}=\frac{1}{\sqrt{L_y}}\varphi_{\lvert n \rvert}(x-X)e^{iyX/l^2}$ that diagonalize the kinetic term in Eq. \ref{2.1.1}, where $\varphi_{\lv n \rv}(x-X)$ are the usual 1D harmonic oscillator (SHO) functions, and $L_y$ is the system size in the $\hat{y}$ direction. Each LL has a macroscopic degeneracy $g=\frac{S}{2\pi\ell^2}$, where $S=L_xL_y$ is the area of the graphene sample. The resulting expansion takes the form
\beq
\begin{aligned}
&\Psi_{\sigma,K}(\bo{r},A)=\sum_{X,n}\frac{\big(\sqrt{2} \big)^{\delta_{n,0}}}{\sqrt{2}}\phi_{\lvert {n} \rvert,X}(\bo{r})c_{nX\sigma K}, \
\Psi_{\sigma,K}(\textbf{r},B)=\sum_{X,n}\frac{sgn(n)}{\sqrt{2}}\phi_{\lvert n \rvert -1,X}(\bo{r})c_{nX\sigma K};\\
&\Psi_{\sigma,K^{\prime}}(\textbf{r},A)=\sum_{X,n}\frac{sgn(n)}{\sqrt{2}}\phi_{\lvert n \rvert -1,X}(\bo{r})c_{nX\sigma K^{\prime}}, \ 
\Psi_{\sigma,K^{\prime}}(\textbf{r},B)=\sum_{X,n}\frac{\big(\sqrt{2} \big)^{\delta_{n,0}}}{\sqrt{2}}\phi_{\lvert n \rvert,X}(\bo{r})c_{nX\sigma K^{\prime}}.\\
\end{aligned} \label{2.1.6} 
\eeq
In these expressions, we define $sgn(n=0)=0$, so that, for the zeroth LL (zLL), the valley pseudospin $K$ ($K^{\prime}$) coincides with the sublattice A (B).

We proceed to form the Hartree-Fock Hamiltonian by pairing the interaction terms in the standard way, resulting in direct (D) and exchange (X) terms. Only the exchange term is kept for the Coulomb interaction; it is assumed that a uniform charged background in the system cancels the direct portion. The HF decomposition thus yields
\beq
\begin{aligned}
\mathcal{H}^{(C)}&\longrightarrow H^{(C)}_{X}\\&=-\sum_{\sigma',\sigma,\tau,\tau'}\sum_{\bo{n},\bo{X}}\mathcal{T}^{(C)}_X(\bo{n};\bo{X})\braket{c^{\dagger}_{n_2X_2\sigma'\tau'}c_{n_4X_4\sigma\tau}}c^{\dagger}_{n_1X_1\sigma\tau}c_{n_3X_3\sigma'\tau'},\\
\mathcal{H}^{(OS)} &\longrightarrow H^{(OS)}_{D}+H^{(OS)}_{X}\\&=\sum_{\sigma'\sigma}\sum_{\bo{n},\bo{X},\bm{\tau}}\bigg[\mathcal{T}^{(OS)}_{D}(\bo{n};\bo{X};\bm{\tau})\braket{c^{\dagger}_{n_2X_2\sigma'\tau_2}c_{n_3X_3\sigma'\tau_3}}c^{\dagger}_{n_1X_1\sigma\tau_1}c_{n_4X_4\sigma\tau_4}\\&\hspace{43pt}-\mathcal{T}^{(OS)}_{X}(\bo{n};\bo{X};\bm{\tau})\braket{c^{\dagger}_{n_2X_2\sigma'\tau_2}c_{n_4X_4\sigma\tau_4}}c^{\dagger}_{n_1X_1\sigma\tau_1}c_{n_3X_3\sigma'\tau_3} \bigg], \\
\text{and} \hspace{50pt} &\\
\mathcal{H}^{(NN)}&\longrightarrow H^{(NN)}_{D}+H^{(NN)}_{X}\\&=\sum_{\sigma'\sigma}\sum_{\bo{n},\bo{X},\bm{\tau}}\bigg[\mathcal{T}^{(NN)}_{D}(\bo{n};\bo{X};\bm{\tau})\braket{c^{\dagger}_{n_2X_2\sigma'\tau_2}c_{n_3X_3\sigma'\tau_3}}c^{\dagger}_{n_1X_1\sigma\tau_1}c_{n_4X_4\sigma\tau_4}\\&\hspace{43pt}-\mathcal{T}^{(NN)}_{X}(\bo{n};\bo{X};\bm{\tau})\braket{c^{\dagger}_{n_2X_2\sigma'\tau_2}c_{n_4X_4\sigma\tau_4}}c^{\dagger}_{n_1X_1\sigma\tau_1}c_{n_3X_3\sigma'\tau_3} \bigg]. \label{2.1.7}
\end{aligned}
\eeq
In the Coulomb exchange,
\beq
\begin{aligned}
\mathcal{T}^{(C)}_X(\bo{n};\bo{X})&\equiv D(\bo{n})\int d^2rd^2r' V^{(C)}(\bo{r}-\bo{r}')\\ 
\times&\bigg [\phi_{\lvert n_1 \rvert, X_1}^{\ast}(\bo{r})\phi_{\lvert n_2 \rvert, X_2}^{\ast}(\bo{r}')\phi_{\lvert n_3 \rvert, X_3}(\bo{r}')\phi_{\lvert n_4 \rvert, X_4}(\bo{r}) \\
&+sgn(n_1n_4)\phi_{\lv n_1 \rv-1, X_1}^{\ast}(\bo{r})\phi_{\lv n_2 \rv, X_2}^{\ast}(\bo{r}')\phi_{\lv n_3 \rv, X_3}(\bo{r}')\phi_{\lv n_4 \rv-1, X_4}(\bo{r})\\
&+sgn(n_2n_3)\phi_{\lv n_1 \rv,X_1}^{\ast}(\bo{r})\phi_{\lv n_2 \rv-1,X_2}^{\ast}(\bo{r}')\phi_{\lv n_3 \rv-1,X_3}(\bo{r}')\phi_{\lv n_4 \rv,X_4}(\bo{r})\\
&+sgn(n_1n_2n_3n_4)\phi_{\lv n_1 \rv-1,X_1}^{\ast}(\bo{r})\phi_{\lv n_2 \rv-1,X_2}^{\ast}(\bo{r}')\phi_{\lv n_3 \rv-1,X_3}(\bo{r}')\phi_{\lv n_4 \rv-1,X_4}(\bo{r}) \bigg], \label{2.1.8}
\end{aligned}
\eeq
where $D(\bo{n})=\prod_{i=1}^4\sqrt{2}^{(\delta_{n_i,0}-1)}$ is a normalization factor. Note the form of these matrix elements is independent of the valley index, a direct consequence of the SU(4) symmetric nature of the (long-range) Coulomb interaction. The absolute value sign on the LL indices, in the SHO wave functions, will be implied for rest of this section.

The expression above involves the integral,
\beq
\widetilde{I}^{(i)}(\bo{n};\bo{X})\equiv \int d^2rd^2r'V^{(i)}(\bo{r}-\bo{r}')\phi_{n_1, X_1}^{\ast}(\bo{r})\phi_{n_2, X_2}^{\ast}(\bo{r}')\phi_{n_3, X_3}(\bo{r}')\phi_{n_4, X_4}(\bo{r}) \label{2.1.9},
\eeq
and Fourier transforming the potential $\big(V^{(i)}(\bo{r})=\int\frac{d^2q}{(2\pi)^2}e^{-i\bo{q}\cdot\bo{r}}\widetilde{V}^{(i)}(q) \big)$, gives
\beq
\widetilde{I}^{(i)}(\bo{n};\bo{X})=\int\frac{d^2q}{(2\pi)^2}\widetilde{V}^{(i)}(q)\braket{n_1X_1\lv e^{-i\bo{q}\cdot\bo{r}} \rv n_4X_4}\braket{n_2X_2\lv e^{i\bo{q}\cdot\bo{r}} \rv n_3X_3}, \label{2.1.10}
\eeq
where,
\beq
\braket{n'X' \lv e^{i\bo{q}\cdot\bo{r}}\rv nX}=e^{iq_x(X+X')/2}F_{n',n}(\bo{q})\delta_{X',X+q_y\ell^2}, \label{2.1.11}
\eeq
with the form factor \cite{MacDonald1991} given by
\beq
F_{n',n}(\bo{q})=\bigg(\frac{n!}{n'!}\bigg)^{1/2}\bigg[ \frac{(-q_y+iq_x)\ell}{\sqrt{2}} \bigg]^{n'-n}e^{-q^2\ell^2/4}L^{n'-n}_n(q^2\ell^2/2) \ \text{for} \ n' \geq n. \label{2.1.12}
\eeq
In Eq. \ref{2.1.12}, $L^{\alpha}_{\beta}(x)$ are the associated Laguerre polynomials, and the form factors have the property that $F_{n,n'}(\bo{q})= \big[ F_{n',n}(-\bo{q}) \big]^{\ast}$.

We impose the condition that $\braket{c^{\dagger}_{nX\sigma\tau}c_{n'X'\sigma'\tau'}}=\delta_{X,X'}\braket{c^{\dagger}_{n\sigma\tau}c_{n'\sigma'\tau'}}$, so that the states are spatially homogenous. This assumption introduces delta functions within the HF Hamiltonian which, in combination with the delta function restrictions and $X$-dependent phase factors in Eq. \ref{2.1.11}, will collapse three of the four guiding center sums. All of the matrix elements needed for the calculation then become independent of the guiding center index, and the integrals relevant to the HF Hamiltonian are then of the form
\beq
I^{(i)}(\bo{n})=\int\frac{d^2q}{(2\pi)^2}\widetilde{V}^{(i)}(q)F_{n_1,n_4}(-\bo{q})F_{n_2,n_3}(\bo{q}). \label{2.1.13}
\eeq
For the Coulomb potential, this becomes
\beq
I^{(C)}(\bo{n})=\frac{\sqrt{2}e^2}{8\pi^2\epsilon_0\ell}J_{min(n_1,n_4),min(n_2,n_3)}^{\lv n_1-n_4 \rv}\delta_{n_1-n_4,n_3-n_2}, \label{2.1.14}
\eeq
where
\beq
J_{n,m}^{\alpha}\equiv \bigg( \frac{n! \ m!}{(\alpha+n)! \ (\alpha+m)!}\bigg)^{\frac{1}{2}}\int_0^{\infty}dx \ x^{2\alpha}e^{-x^2}L_n^{\alpha}(x^2)L^{\alpha}_m(x^2). \label{2.1.15}
\eeq
This integral is solved numerically using recursive properties of the associated Laguerre polynomials to create recursion releations for $J^{\alpha}_{n,m}$, allowing efficient computation for large values of the parameters.
Inserting the contact potential into Eq. \ref{2.1.13} yields simpler results,
\beq
\begin{aligned}
I^{(OS)}_{X}(\bo{n})=\frac{V_0a_c}{8\pi\ell^2}\delta_{n_1,n_3}\delta_{n_2,n_4} \ , \ I^{(NN)}_{X}(\bo{n})=\frac{3V_1a_c}{16\pi\ell^2}\delta_{n_1,n_3}\delta_{n_2,n_4}, \label{2.1.16}
\end{aligned}
\eeq
and
\beq
I^{(OS)}_{D}(\bo{n})=\frac{V_0a_c}{8\pi\ell^2}\delta_{n_1,n_4}\delta_{n_2,n_3} \ , \ I^{(NN)}_{D}(\bo{n})=\frac{3V_1a_c}{16\pi\ell^2}\delta_{n_1,n_4}\delta_{n_2,n_3}. \label{2.1.17}
\eeq
The $X$ ($D$) label differentiates between $I$'s appearing in the exchange (direct) terms.

For the Coulomb interaction,
\beq
\begin{aligned}
\mathcal{T}_X^{(C)}(\bo{n})=&D(\bo{n})\bigg[I^{(C)}(\bo{n})+sgn(n_1n_4)I^{(C)}(n_1-1,n_2,n_3,n_4-1) \\
&\hspace{25pt}+sgn(n_2n_3)I^{(C)}(n_1,n_2-1,n_3-1,n_4)+sgn(n_1n_2n_3n_4)I^{(C)}(\bo{n}-1)\bigg],
\end{aligned} \label{2.1.18}
\eeq
where $(\bo{n}-1)\equiv (n_1-1,n_2-1,n_3-1,n_4-1)$. For the OS and NN interactions there are six combinations of valley indices which give non-vanishing results. They are:
\\
(i) $\tau_1=\tau_2=\tau_3=\tau_4=K$, for which
\beq
\begin{aligned}
\mathcal{T}_j^{(OS)}(\bo{n};K,K,K,K)&=D(\bo{n})\big[I_j^{(OS)}(\bo{n})+sgn(n_1n_2n_3n_4)I_j^{(OS)}(\bo{n}-1)\big],\\
\mathcal{T}_j^{(NN)}(\bo{n};K,K,K,K)&=D(\bo{n})\big[sgn(n_2n_3)I_j^{(NN)}(n_1,n_2-1,n_3-1,n_4)\\&\hspace{100pt}+sgn(n_1n_4)I_j^{(NN)}(n_1-1,n_2,n_3,n_4-1)\big];
\end{aligned}
\eeq
(ii) $\tau_1=\tau_2=\tau_3=\tau_4=K^{\prime}$, for which
\beq
\begin{aligned}
\mathcal{T}_j^{(OS)}(\bo{n};K^{\prime},K^{\prime},K^{\prime},K^{\prime})&=\mathcal{T}_j^{(OS)}(\bo{n};K,K,K,K),\\
\mathcal{T}_j^{(NN)}(\bo{n};K^{\prime},K^{\prime},K^{\prime},K^{\prime})&=\mathcal{T}_j^{(NN)}(\bo{n};K,K,K,K);
\end{aligned}
\eeq
(iii) $\tau_1=\tau_4=K^{\prime} \ , \ \tau_2=\tau_3=K$, for which
\beq
\begin{aligned}
\mathcal{T}_j^{(OS)}(\bo{n};K^{\prime},K,K,K^{\prime})&=\mathcal{T}_j^{(NN)}(\bo{n};K,K,K,K),\\
\mathcal{T}_j^{(NN)}(\bo{n};K^{\prime},K,K,K^{\prime})&=\mathcal{T}_j^{(OS)}(\bo{n};K,K,K,K);
\end{aligned}
\eeq
(iv) $\tau_1=\tau_4=K \ , \ \tau_2=\tau_3=K^{\prime}$, for which
\beq
\begin{aligned}
\mathcal{T}_j^{(OS)}(\bo{n};K,K^{\prime},K^{\prime},K)&=\mathcal{T}_j^{(NN)}(\bo{n};K,K,K,K),\\
\mathcal{T}_j^{(NN)}(\bo{n};K,K^{\prime},K^{\prime},K)&=\mathcal{T}_j^{(OS)}(\bo{n};K,K,K,K);
\end{aligned}
\eeq
(v) $\tau_1=\tau_3=K^{\prime} \ , \ \tau_2=\tau_4=K$, for which
\beq
\begin{aligned}
\mathcal{T}_j^{(OS)}(\bo{n};K^{\prime},K,K^{\prime},K)&=D(\bo{n})\big[sgn(n_2n_4)I_j^{(OS)}(n_1,n_2-1,n_3,n_4-1)\\&\hspace{100pt}+sgn(n_1n_3)I_j^{(OS)}(n_1-1,n_2,n_3-1,n_4)\big], \\
\mathcal{T}_j^{(NN)}(\bo{n};K^{\prime},K,K^{\prime},K)&=D(\bo{n})\big[sgn(n_3n_4)I_j^{(NN)}(n_1,n_2,n_3-1,n_4-1)\\&\hspace{100pt}+sgn(n_1n_2)I_j^{(NN)}(n_1-1,n_2-1,n_3,n_4)\big];
\end{aligned}
\eeq
and (vi) $\tau_1=\tau_3=K \ , \ \tau_2=\tau_4=K^{\prime}$, for which
\beq
\begin{aligned}
\mathcal{T}_j^{(OS)}(\bo{n};K,K^{\prime},K,K^{\prime})&=\mathcal{T}_j^{(OS)}(\bo{n};K^{\prime},K,K^{\prime},K), \\
\mathcal{T}_j^{(NN)}(\bo{n};K,K^{\prime},K,K^{\prime})&=\mathcal{T}_j^{(NN)}(\bo{n};K^{\prime},K,K^{\prime},K).
\end{aligned}
\eeq
\subsection{Dirac sea}
\begin{figure}[h]
\centering
\captionbox
{Schematic diagram of the LL structure. The inactive levels are assumed to be filled, and cannot admix with other levels.\label{Figure1}}
{\includegraphics[width=0.4\linewidth]{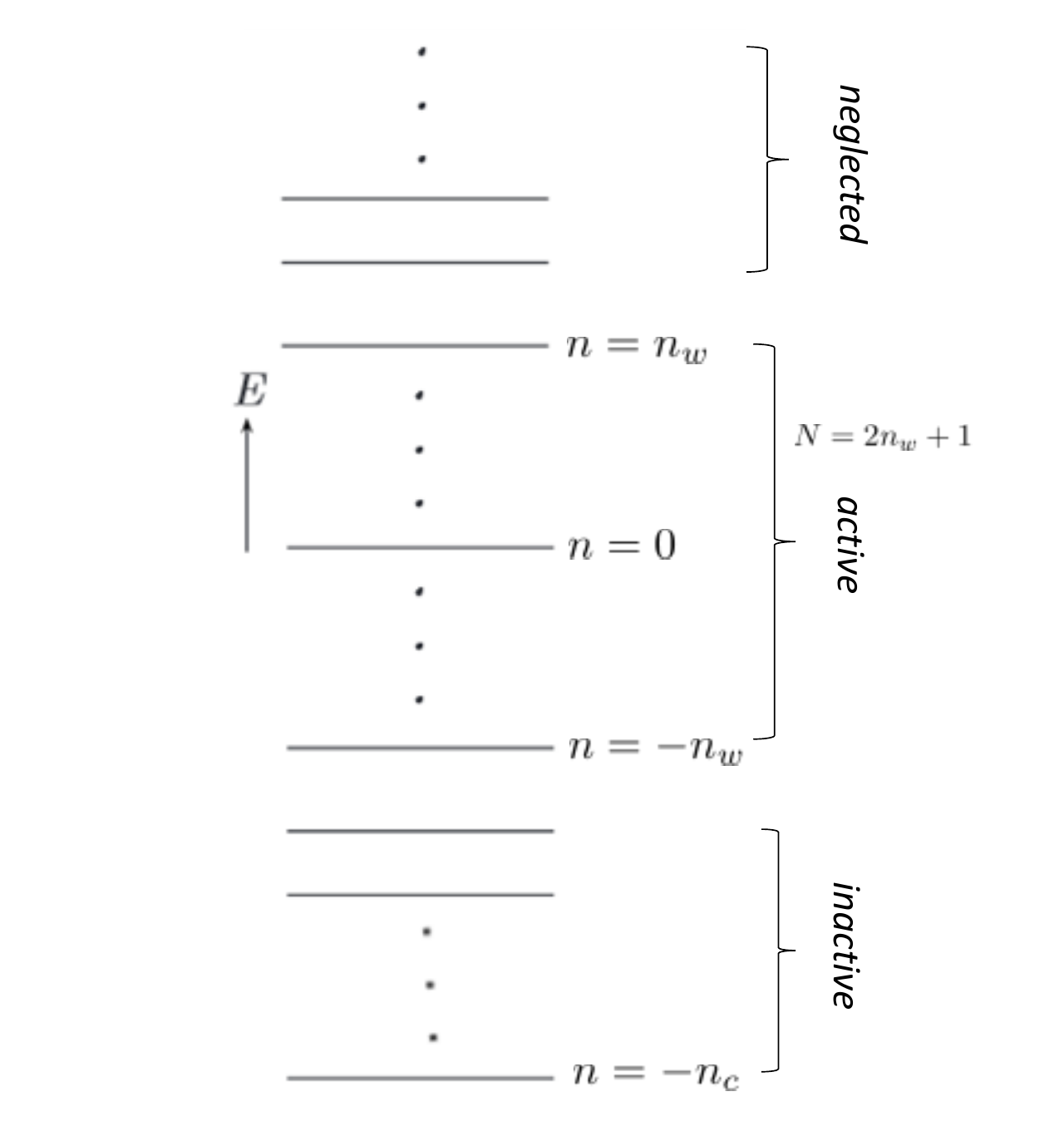}}
\end{figure}

One challenge in carrying out this calculation is the large number of negative energy LL's, which are filled with electrons, and, as we shall see, can support quantitatively significant LL mixing, even very far from the Fermi energy. In practice one may only retain a finite number of these LL's when optimizing the HF state. To proceed, we will assume the negative energy levels that are not actively retained \cite{Jianhui2008} are filled, and ignore any LL mixing they may host. We call these lowest filled ``inactive" levels the Dirac sea, and their presence is not completely inert; they present an effective potential to the remaining ``active" LL's, which enters as a single-particle term in the HF Hamiltonian. The organization of the LL's in the calculation is illustrated in Fig. \ref{Figure1}.
The lowest LL in the sea is determined by assuming that there is only one electron per carbon atom, which is the density of electrons in the $p_z$ orbitals of graphene. The index for this cut-off thus obeys
\beq
-n_c=-\frac{2\pi\ell^2}{\sqrt{3}\tilde{a}^2}\propto\frac{1}{B_\perp}, \label{2.2.1}.
\eeq
where $\tilde{a}=0.246$nm is the triangular Bravais lattice constant. The active window of LLs is centered about $n=0$, preserving particle-hole symmetry. We denote the lowest LL index in the active window as $-n_w$, so that the total number of LLs in the active window is $N\equiv(2n_w+1)$. LL's in the range $-n_c \leq n < -n_w$ thus belong to the sea, whereas a LL in the range $-n_w \leq n \leq n_w$ is included in the active window. LL's with $n > n_w$ are neglected in this calculation (see Fig. \ref{Figure1}).

To avoid confusion, the absolute value signs on the LL indices will be included in this section. In the Dirac sea, it is assumed that the density matrix has the (structureless) form 
\beq
\braket{c^{\dagger}_{n'\sigma'\tau'}c_{n\sigma\tau}}=\delta_{n,n'}\delta_{\tau,\tau'}\delta_{\sigma,\sigma'},\label{2.2.2}
\eeq
with $-n_c \leq n< -n_w$ and $-n_c \leq n'< -n_w$. The Coulomb interaction then induces a term in the HF Hamiltonian of the form
\beq
\mathcal{H}_{X,sea}^{(C)}=-\sum_{n_1,n_3=-n_w}^{n_w}\bigg[\sum_{n=-n_c}^{-n_w-1} \mathcal{T}^{(C)}_{X,sea}(\bo{n}_{24}) \bigg]\sum_{\sigma,\tau,X}c^{\dagger}_{n_1X\sigma\tau}c_{n_3X\sigma\tau},\label{2.2.3}
\eeq
where $\bo{n}_{24}\equiv(n_1,n,n_3,n)$. LL indices within the Dirac sea are always negative, and larger in magnitude than the LL's in the active window ($\lv n \rv > \lv n_1 \rv,\lv n_3 \rv$). Using Eqs. [\ref{2.1.13}] and [\ref{2.1.17}], and noting that $J^{\alpha}_{n,m}=J^{\alpha}_{m,n}$, we can then write
\beq
\begin{aligned}
\sum_{n=-n_c}^{-n_w-1}& \mathcal{T}_{X,sea}^{(C)}(\bo{n}_{24}) \\
=&\sum_{n=-n_c}^{-n_w-1}D(\bo{n}_{24})\bigg[ I^{(C)}_X(\lv n_1 \rv,\lv n \rv, \lv n_3\rv, \lv n \rv) -sgn(n_1)I^{(C)}_X(\lv n_1 \rv-1,\lv n \rv, \lv n_3 \rv, \lv n \rv -1) \\
&-sgn(n_3)I^{(C)}_X(\lv n_1 \rv, \lv n \rv -1, \lv n_3 \rv -1, \lv n \rv)\\
&+sgn(n_1n_3)I^{(C)}_X(\lv n_1 \rv -1, \lv n \rv -1, \lv n_3 \rv -1, \lv n \rv -1) \bigg]\\
=&\frac{\sqrt{2}e^2}{8\pi^2\epsilon_0\ell}\sum_{n=-n_c}^{-n_w-1}\delta_{\lv n_1 \rv,\lv n_3 \rv}A(n_1,n_3,n),\label{2.2.4}
\end{aligned}
\eeq
with
\beq
  A(n_1,n_3,n) \equiv \left\{\def\arraystretch{1.2}%
  \begin{array}{@{}c@{\quad}l@{}}
  	\frac{1}{4}\bigg[ J_{\lv n' \rv,\lv n' \rv}^{\lv n \rv - \lv n' \rv} -J_{\lv n' \rv-1,\lv n' \rv-1}^{\lv n \rv - \lv n' \rv} \bigg] & \text{if $n_1n_3 < 0$} \ \ (\lv n' \rv=\lv n_1 \rv=\lv n_3 \rv) \\ \\
    \frac{1}{2}J^{\lv n \rv}_{0,0} & \text{if $n_1=n_3=0$}\\ \\
    \frac{1}{4}\bigg[J_{\lv n' \rv,\lv n' \rv}^{\lv n \rv - \lv n' \rv}-sgn(n') \ 2 J_{\lv n' \rv-1,\lv n' \rv}^{\lv n \rv - \lv n' \rv} -J_{\lv n' \rv-1,\lv n' \rv-1}^{\lv n \rv - \lv n' \rv}\bigg] & \text{if $n_1n_3>0$} \ \ (n'=n_1=n_3).
  \end{array}\right.\label{2.2.5}
\eeq
The term in $A(n_1,n_3,n)$ with $n_1n_3<0$ adds a small, but non-zero, contribution to the LL mixing within the active window.

The short-range interaction effects of the Dirac sea work in much the same way. Starting with the HF components for each interaction,
\beq
\begin{aligned}
\mathcal{H}^{(OS)}_D+\mathcal{H}^{(NN)}_D+&\mathcal{H}^{(OS)}_X+\mathcal{H}^{(NN)}_X\\&=\sum_{\bo{n},\bm{\tau}}\bigg[\mathcal{T}^{(OS)}_{D}(\bo{n};\bm{\tau})+\mathcal{T}^{(NN)}_{D}(\bo{n};\bm{\tau}) \bigg]\sum_{\sigma'}\braket{c^{\dagger}_{n_2\sigma'\tau_2}c_{n_3\sigma'\tau_3}}\sum_{X,\sigma}c^{\dagger}_{n_1X\sigma\tau_1}c_{n_4X\sigma\tau_4}\\
&-\sum_{\sigma,\sigma'}\sum_{\bo{n},\bm{\tau}}\bigg[\mathcal{T}^{(OS)}_{X}(\bo{n};\bm{\tau})+\mathcal{T}^{(NN)}_{X}(\bo{n};\bm{\tau}) \bigg]\braket{c^{\dagger}_{n_2\sigma'\tau_2}c_{n_4\sigma\tau_4}}\sum_Xc^{\dagger}_{n_1X\sigma\tau_1}c_{n_3X\sigma'\tau_3}, \label{2.2.6}
\end{aligned}
\eeq
and using Eq. \ref{2.2.2} for the density matrix in the sea, the exchange terms become
\beq
\begin{aligned}
\mathcal{H}^{(OS)}_{X,sea}&+\mathcal{H}^{(NN)}_{X,sea}=-\sum_{\tau_1,\tau_3,\tau} \ \sum_{n_1,n_3}\sum_{n=-n_c}^{-n_w-1}\bigg[\mathcal{T}^{(OS)}_{X}(\bo{n}_{24};\bm{\tau}_{24})+\mathcal{T}^{(NN)}_{X}(\bo{n}_{24};\bm{\tau}_{24}) \bigg]\sum_{X,\sigma}c^{\dagger}_{n_1X\sigma\tau_1}c_{n_3X\sigma\tau_3}.
\end{aligned}\label{2.2.7}
\eeq
\\There are four possible combinations of valley indices (cases i., ii., v., and vi. above) which satisfy $\tau_2=\tau_4$, and, in each case, the constraint $\tau_1=\tau_3$ must also be satisfied, so that $f(\bm{\tau})=1$. Furthermore, the $\mathcal{T}$ matrix elements belonging to those cases are independent of whether $\tau_1=\tau_3=K$ or $\tau_1=\tau_3=K^{\prime}$. Lastly, the NN exchange term vanishes within the sum since $\mathcal{T}^{(NN)}_{X}(\bo{n}_{24};\bm{\tau}_{24}) \propto \delta_{\lv n \rv -1,\lv n \rv}=0$. Eq. \ref{2.2.7} then reduces to
\beq
\begin{aligned}
\mathcal{H}^{(OS)}_{X,sea}&+\mathcal{H}^{(NN)}_{X,sea}\\&=-\frac{V_oa_c}{4\pi\ell^2}N_{sea}\sum_{n_1,n_3=-n_w}^{n_w}\frac{(\sqrt{2})^{\delta_{n_1,0}+\delta_{n_3,0}}}{4}\big[1+sgn(n_1n_3)\big]\delta_{\lv n_1 \rv,\lv n_3 \rv}\sum_{X,\sigma,\tau}c^{\dagger}_{n_1X\sigma\tau}c_{n_3X\sigma\tau},
\end{aligned}\label{2.2.8}
\eeq
where $N_{sea}=n_c-n_w$ is the total number of LLs in the Dirac sea. 

The direct components for the interactions are handled in a similar way. We write 
\beq
\begin{aligned}
\mathcal{H}^{(OS)}_{D,sea}&+\mathcal{H}^{(NN)}_{D,sea}\\&=2\sum_{\tau_1,\tau_4,\tau} \ \sum_{n_1,n_4}\sum_{n=-n_c}^{-n_w-1}\bigg[\mathcal{T}^{(OS)}_{D}(\bo{n}_{23};\bm{\tau}_{23})+\mathcal{T}^{(NN)}_{D}(\bo{n}_{23};\bm{\tau}_{23})\bigg]\sum_{X,\sigma}c^{\dagger}_{n_1X\sigma\tau_1}c_{n_4X\sigma\tau_4},\label{2.2.9}
\end{aligned}
\eeq
with the additional factor of 2 coming from summing over spin. We then obtain
\beq
\begin{aligned}
\mathcal{H}^{(OS)}_{D,sea}&+\mathcal{H}^{(NN)}_{D,sea}\\&=\frac{a_c}{2\pi\ell^2}\big(V_0+\frac{3V_1}{2}\big)N_{sea}\sum_{n_1,n_4=-n_w}^{n_w}\frac{(\sqrt{2})^{\delta_{n_1,0}+\delta_{n_4,0}}}{4}\big[1+sgn(n_1n_4)\big]\delta_{\lv n_1 \rv, \lv n_4 \rv}\sum_{X,\sigma,\tau}c^{\dagger}_{n_1X\sigma\tau}c_{n_4X\sigma\tau}.\label{2.2.10}
\end{aligned}
\eeq
Finally, relabeling indices ($n_4 \leftrightarrow n_3$) for the direct terms and combining with the exchange term yields
\beq
\begin{aligned}
\mathcal{H}^{(OS)}_{D,sea}&+\mathcal{H}^{(NN)}_{D,sea}+\mathcal{H}^{(OS)}_{X,sea}+\mathcal{H}^{(NN)}_{X,sea}\\
&=\frac{(V_0+3V_1)a_c}{4\pi\ell^2}N_{sea}\sum_{n_1,n_3=-n_w}^{n_w}\frac{(\sqrt{2})^{\delta_{n_1,0}+\delta_{n_3,0}}}{4}\bigg(1+sgn(n_1n_3) \bigg)\delta_{\lv n_1 \rv,\lv n_3 \rv}\sum_{X,\sigma,\tau}c^{\dagger}_{n_1X\sigma\tau}c_{n_3X\sigma\tau}\\
&=\frac{1}{2}\frac{(V_0+3V_1)a_c}{4\pi\ell^2}N_{sea}\sum_{n,X,\sigma,\tau}c^{\dagger}_{nX\sigma\tau}c_{nX\sigma\tau}.\label{2.2.11}
\end{aligned}
\eeq
Thus the short-range interactions yield a term in the HF Hamiltonian which is diagonal in all the state indices. Therefore, the states in the active window are unaffected by short-range interactions with electrons in the filled sea; the only non-trivial contribution comes from the Coulomb interaction. With this contribution, our HF Hamiltonian may be written in the form
\beq
\begin{aligned}
&H_{HF}=\\
&\sum_{n_1,n_3=-n_w}^{n_w}\sum_{\sigma}\bigg[sgn(n_1)\frac{\hbar V_F}{\ell}\sqrt{2\lv n_1 \rv}\delta_{n_1,n_3}-\frac{\sigma g^{\ast}\mu_B}{2}\delta_{n_1,n_3}-\mathcal{T}_{X,sea}^{(C)}(n_1,n_3)\bigg]\sum_{X,\tau}c^{\dagger}_{n_1X\sigma\tau}c_{n_3X\sigma\tau}\\
&\hspace{107pt}+\sum_{\bo{n},\bm{\tau}}\bigg[\mathcal{T}^{(OS)}_{D}(\bo{n};\bm{\tau})+\mathcal{T}_{D}^{(NN)}(\bo{n};\bm{\tau})\bigg]\sum_{\sigma'}\braket{c^{\dagger}_{n_1\sigma'\tau}c_{n_3\sigma'\tau_3}}\sum_{X,\sigma}c^{\dagger}_{n_1X\sigma\tau_1}c_{n_4X\sigma\tau_4}\\
&-\sum_{\bo{n},\bm{\tau}}\bigg[\bigg(\mathcal{T}^{(OS)}_{X}(\bo{n};\bm{\tau})+\mathcal{T}^{(NN)}_{X}(\bo{n};\bm{\tau})\bigg)+\delta_{\tau_2,\tau_3}\delta_{\tau_1,\tau_4}\mathcal{T}_X^{(C)}(\bo{n}) \bigg]\sum_{\sigma,\sigma'}\braket{c^{\dagger}_{n_2\sigma'\tau_2}c_{n_4\sigma\tau_4}}\sum_Xc^{\dagger}_{n_1X\sigma\tau_1}c_{c_3X\sigma'\tau_3},
\end{aligned}\label{effectivehamiltonian}
\eeq
where each LL index above now includes only the LL's within the active window, and we have dropped a constant term.
\section{Hartree-Fock States}
The HF approximation requires an initial guess for the state of the system, which may be expressed via the density matrix. Following Ref. \onlinecite{Kharitonov2012}, we consider here the ferromagnet (FM), charge density (CDW) ordered, Kekul\'{e} dimerized (KD), and canted antiferromagnetic (CAF) states. The simplest way to describe these states is by projecting the system into the zLL, where we can then identify a particular sublattice with a single valley index. At zero filling factor, the chemical potential is within the $n=0$ LL's. Because of the discrete index structure (spin and valley), there are four quantum labels which are needed to specify a state ($\uparrow\downarrow \otimes \ KK^{\prime}$) in the zLL. Another property of filling factor $\nu=0$ is that there are on average two electrons per guiding center index in the zLL. Therefore the trace of the density matrix for a particular state, projected into the zLL, must be 2. How these two electrons occupy the zLL defines the state of the system.

The (FM) state has the two electrons spin polarized along the direction of the total magnetic field ($+\hat{z}$), forcing the two electrons to occupy different sublattices, or, equivalently, opposite valleys. A choice for the density matrix representation of the FM state is
\beq
\braket{c^{\dagger}_{0\sigma\tau}c_{0\sigma'\tau'}}_{FM}=\delta_{\sigma,\uparrow}\delta_{\sigma',\uparrow}\delta_{\tau,\tau'}. \label{3.1.1}
\eeq
Note that there are only two non-zero matrix elements here, and that $Tr[\braket{c^{\dagger}_{0\sigma\tau}c_{0\sigma'\tau'}}_{FM}]=2$.

The charge density wave (CDW) has the property that both electrons in the zLL occupy the same sublattice, but with oppositely polarized spins; the state is a spin singlet. Occupying either sublattice leads to a groundstate with the same energy, so that, in this case, there is a broken $Z_2$ symmetry. One choice for the density matrix representation of the CDW state is
\beq
\braket{c^{\dagger}_{0\sigma\tau}c_{0\sigma'\tau'}}_{CDW}=\delta_{\sigma,\sigma'}\delta_{\tau,K}\delta_{\tau',K}. \label{3.1.2}
\eeq
The CDW parallels in some ways the FM state, with the roles of valley and spin reversed. Much like the FM density matrix, the CDW has only two non-zero matrix elements.

Another possible phase is the KD state. The main feature of the KD state is that the single particle wave functions for the two electrons have equal weight in the $K$ and $K^{\prime}$ valley points; the vector representing the valley degree of freedom lies on the equator of the Bloch sphere, breaking a valley U(1) symmetry in the HF Hamiltonian. This state is also a spin singlet, and a possible matrix representation is
\beq
\braket{c^{\dagger}_{0\sigma\tau}c_{0\sigma'\tau'}}_{KD}=\frac{1}{2}\delta_{\sigma,\sigma'}, \label{3.1.3}
\eeq
in which the valley pseudospin points along the $\hat{x}$-direction.

When the Zeeman field is neglected, one possible state which can occur is the anti-ferromagnet (AFM) with density matrix, for example,
\beq
\braket{c^{\dagger}_{0\sigma\tau}c_{0\sigma'\tau'}}_{AFM}=\delta_{\sigma,\uparrow}\delta_{\sigma',\uparrow}\delta_{\tau,K}\delta_{\tau',K}+\delta_{\sigma,\downarrow}\delta_{\sigma',\downarrow}\delta_{\tau,K^{\prime}}\delta_{\tau',K^{\prime}}. \label{3.1.4}
\eeq 
With a non-zero Zeeman field, the AFM is modified into a CAF. For the CAF state the spins of the electrons become partially polarized as they cant in the direction of the applied field, while the in-plane components of the electron spins remain anti-parallel. The CAF breaks a U(1) symmetry in the HF Hamiltonian with respect to rotations of the in-plane spin components about the Zeeman field direction. The density matrix representation for this state can be described using two continuous parameters, $\epsilon$ and $\Delta$, which are, in general, functions of the system parameters. Our choice of the density matrix for the CAF state has the form
\beq
\begin{aligned}
\braket{c^{\dagger}_{0\sigma,\tau}c_{0\sigma',\tau'}}_{CAFM}=&\bigg[\frac{1}{2}(\mathds{1})_{\sigma,\sigma'}+\epsilon \ (\sigma_z)_{\sigma,\sigma'}+\Delta \ (\sigma_x)_{\sigma,\sigma'} \bigg]\delta_{\tau,K}\delta_{\tau',K}\\+&\bigg[\frac{1}{2}(\mathds{1})_{\sigma,\sigma'}+\epsilon \ (\sigma_z)_{\sigma,\sigma'}-\Delta \ (\sigma_x)_{\sigma,\sigma'} \bigg]\delta_{\tau,K^{\prime}}\delta_{\tau',K^{\prime}}, \label{3.1.5}
\end{aligned}
\eeq
where $\sigma_i$ are the usual Pauli matrices for the spin degree of freedom. Note that if we take $\epsilon=1/2$ and $\Delta=0$, we recover the density matrix for the FM. If $\epsilon=0$ and $\Delta=1/2$, the electron occupying the A sublattice ($\tau=\tau'=K$) is spin polarized entirely in the $+\hat{x}$ direction, while the electron on sublattice B ($\tau=\tau'=K^{\prime}$) is spin polarized entirely in the $-\hat{x}$ direction. For initial guesses we used $\epsilon=0$ and $\Delta=1/2$ in the above density matrix as the seed for the CAF in what follows. Upon iteration, if the CAF was a stable solution in a particular region of the parameter space, then the values for $\epsilon$ and $\Delta$ would also change, but they always remained within the intervals $0 \leq\epsilon<1/2$ and $0<\Delta \leq 1/2$.

With these possible forms for the zLL, our initial guesses for the full density matrix in the active window were assumed to have the form
\beq
   \braket{c^{\dagger}_{n\sigma\tau}c_{n'\sigma'\tau'}}_j= \left\{\def\arraystretch{1.2}%
  \begin{array}{@{}c@{\quad}l@{}}
  	\delta_{nn'}\delta_{\sigma\sigma'}\delta_{\tau\tau'} & \text{for \ $-n_w < n < 0$} \\ \\
    \braket{c^{\dagger}_{0\sigma\tau}c_{0\sigma'\tau'}}_j & \text{for \ \ $n=0$} \\ \\
    0 & \text{for \ \ $0 < n < n_w$}.\\
  \end{array}\right. \label{3.2.1}
\eeq
\section{HF groundstates and numerical results}
In the numerical results we report below we focused for concreteness on a perpendicular field of magnitude $B_{\perp}=15T$. Our numerical analysis proceeded by adopting initial guesses for the density matrix of the form in Eq. \ref{3.2.1}, forming the HF Hamiltonian, diagonalizing this, and using the results to obtain a new density matrix. The result was considered converged if $\chi$ $\leq$ $10^{-10}$, where
\beq
\chi \equiv \frac{\sum_{n,n'}\sum_{\sigma,\sigma'}\sum_{\tau,\tau'}\big\lv \rho_{nn'\sigma\sigma'\tau\tau'}^i-\rho_{nn'\sigma\sigma'\tau\tau'}^{i-1} \big\rv}{\sum_{n,n'}\sum_{\sigma,\sigma'}\sum_{\tau,\tau'}\big[\big(\rho_{nn'\sigma\sigma'\tau\tau'}^i \big)^{\ast}\rho_{nn'\sigma\sigma'\tau\tau'}^i \big]^{1/2}}, \label{stopping criterion}
\eeq
with $\rho_{nn'\sigma\sigma'\tau\tau'} \equiv \braket{c^{\dagger}_{n\sigma\tau}c_{n'\sigma'\tau'}}$, and $i$ above labels the iteration. In several cases, we tested the stability of our solutions by adding (small) random additions to the converged density matrix, and used this as a seed for the HF algorithm. In all cases, we found this brought the result back to the previously converged solution.

There were two generally recurring properties of the converged density matrix. The first is that, when our HF algorithm converged, the structure of the zLL would be one of the forms in Eqs. [\ref{3.1.1}] - [\ref{3.1.5}], depending on the initial state. The second important feature is their LL structure. Although we would start with an initial guess which was diagonal in LL index, the program self-consistently generated a groundstate for which the matrix elements $\braket{c^{\dagger}_1c_2}\propto\delta_{\lv n_1 \rv,\lv n_2 \rv}$ were always much larger than the others. Thus, the LLM we find in our states is dominated by mixing between states which are particle-hole partners. Interestingly, this relatively simple form is consistent with the type of trial states considered in Ref. \onlinecite{Shankar2016}.
\subsection{$V_0 - V_1$ phase diagrams}
We begin by presenting phase diagrams for different short-range interaction strengths, for several different active window sizes. To construct these phase diagrams, all four states, starting with the form of Eq. \ref{3.2.1}, were used as initial guesses in the HF algorithm over a range of system parameters ($V_0,V_1$). The total energies of the converged states were then compared. Whichever converged state had the lowest energy represented the HF groundstate.

\begin{figure}[h!]
\centering
\captionbox
{Phase diagram for $N=1$ with small, but non-zero, Zeeman field ($\xi_z << 1$). The two phases appearing in this diagram are the CDW (red) and FM (blue).\label{Figure2}}
{\includegraphics[width=0.5\linewidth]{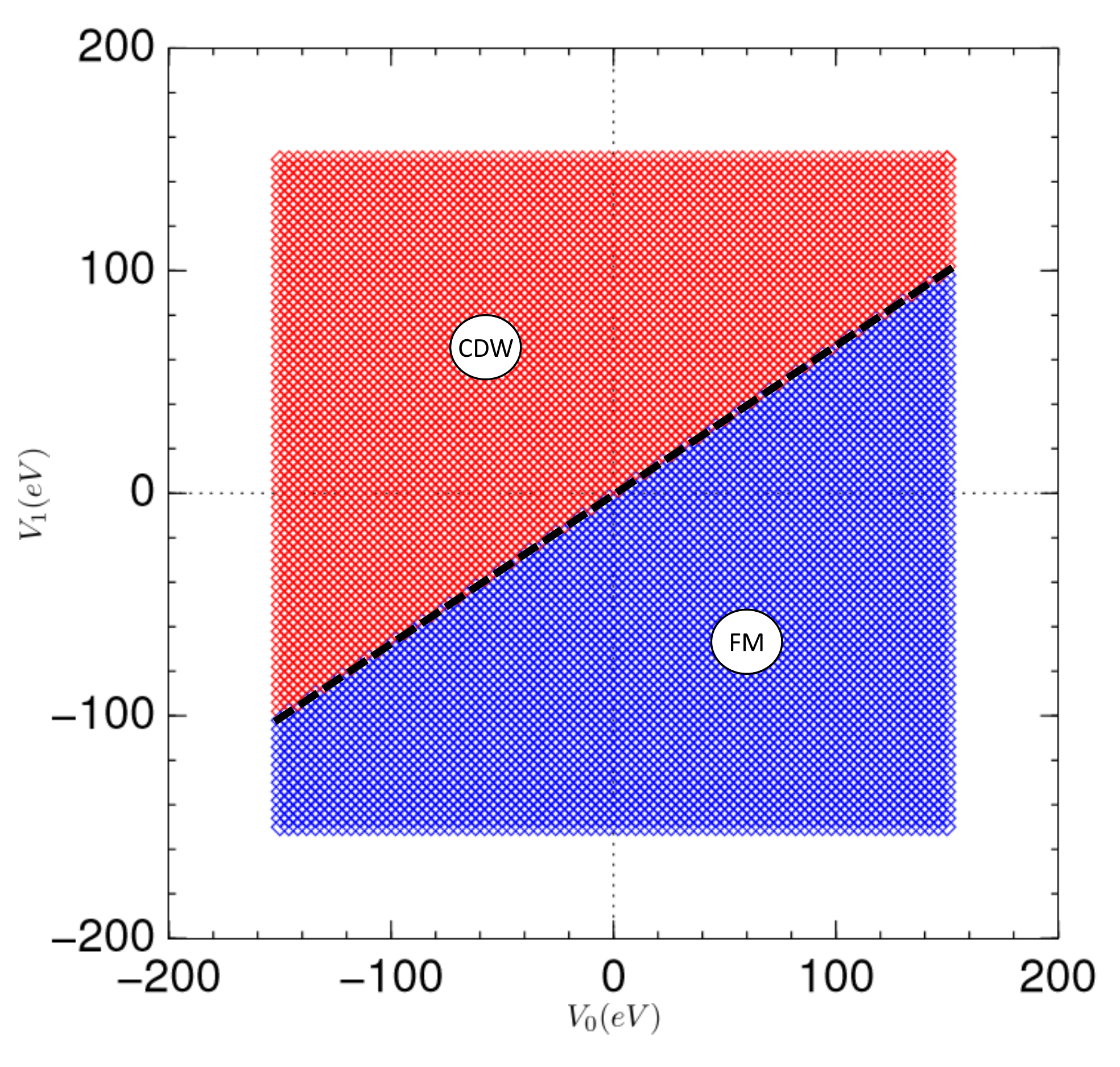}}
\end{figure}

\begin{figure}[h!]
\centering
\captionbox
{Phase diagram for $N=1$ with a larger Zeeman field: $B_T=\xi_zB_{\perp}=30T$.\label{Figure3}}
{\includegraphics[width=0.5\linewidth]{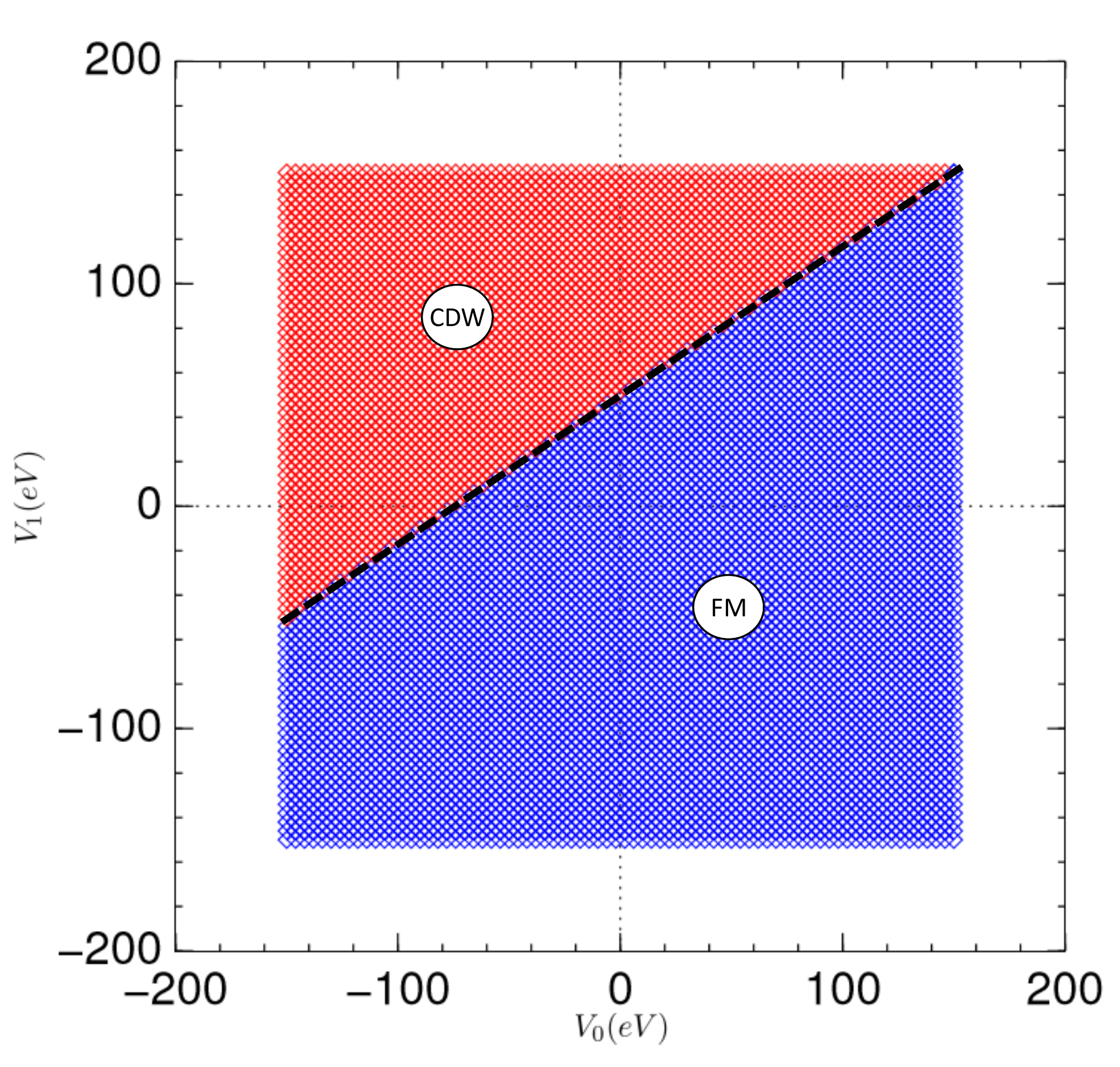}}
\end{figure}

Figure \ref{Figure2} shows the phase diagram for an active window of $N=1$ (keeping only the zLL), in the presence of an infinitesimal Zeeman field ($\xi_z<<1$). The only two phases which appear in this diagram are the CDW (red) and the FM (blue), and the phase boundary separating the two phases represents a first order transition. Notice that, when the OS and NN interactions are zero ($V_0=0,V_1=0$), both phases have the same energy.
Neglecting the NN interaction, and looking at $V_0<0$, the OS interaction is attractive, so it is favorable for the two electrons to occupy the same sublattice, giving rise to the CDW in this portion of the phase diagram. For $V_0>0$, the OS interaction is repulsive, making it more favorable for the electrons to occupy different sublattices. Figure \ref{Figure3} shows the $N=1$ phase diagram for a larger Zeeman field strength. The FM now takes up more area in the parameter space, as should be expected. For the $N=1$ case, in the complete absence of Zeeman field, the AFM and FM states shared the same energy, however, for any non-zero valued field, the FM was always lower in energy than the CAF. The KD phase, although a stable solution to the HF equations for some parameter range, was never the lowest energy solution.

\begin{figure}[h!]
\centering
\captionbox
{$N=3$ phase diagram with no Zeeman field. The FM (blue), AFM (green), and CDW (red) all share the same energy at the origin. Note that the FM is still a stable solution in the region where the AFM now appears.\label{Figure4}}
{\includegraphics[width=0.5\linewidth]{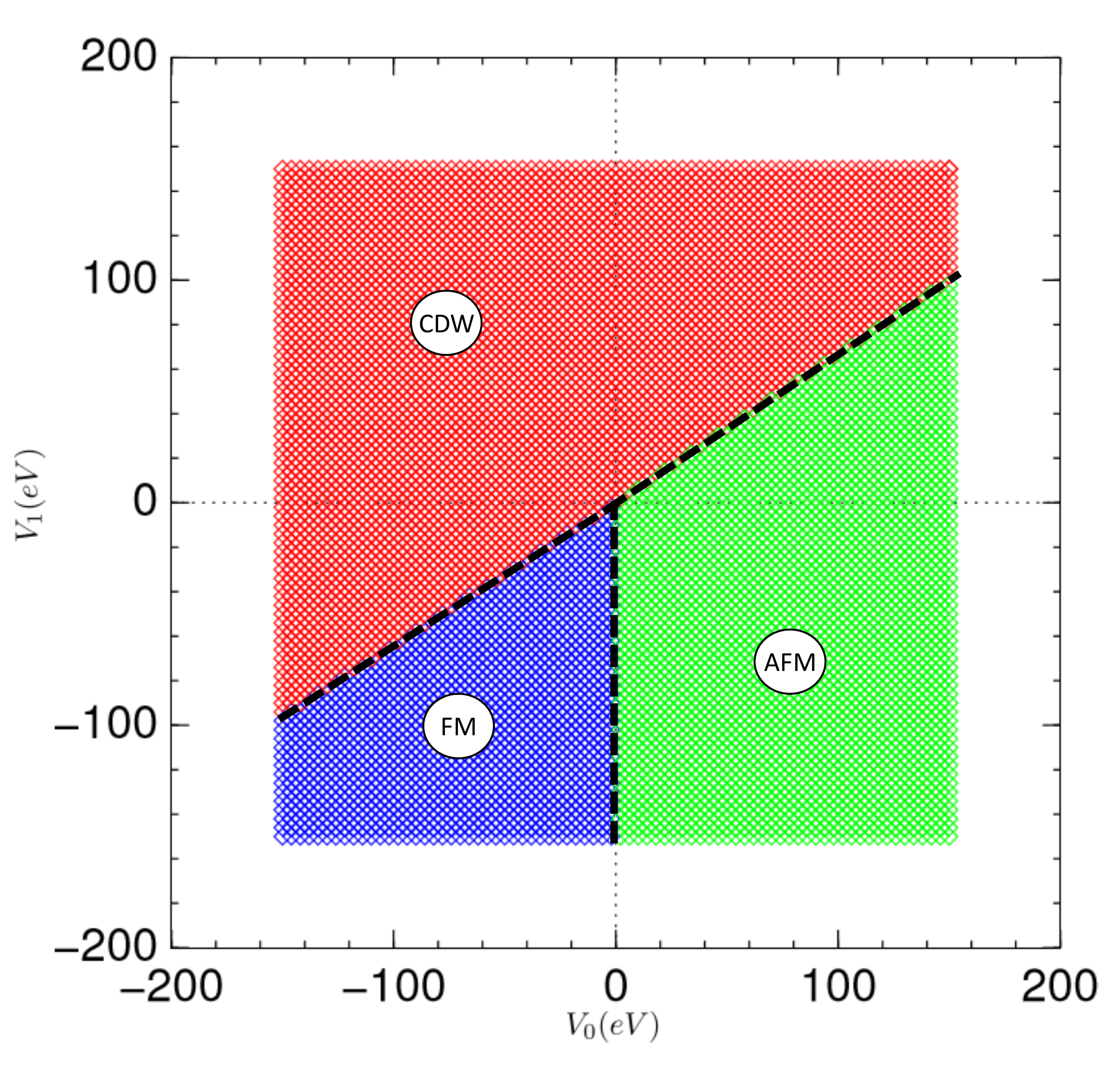}}
\end{figure}

\begin{figure}[h]
\centering
  \begin{subfigure}[b]{0.4\textwidth}
    \includegraphics[width=\textwidth]{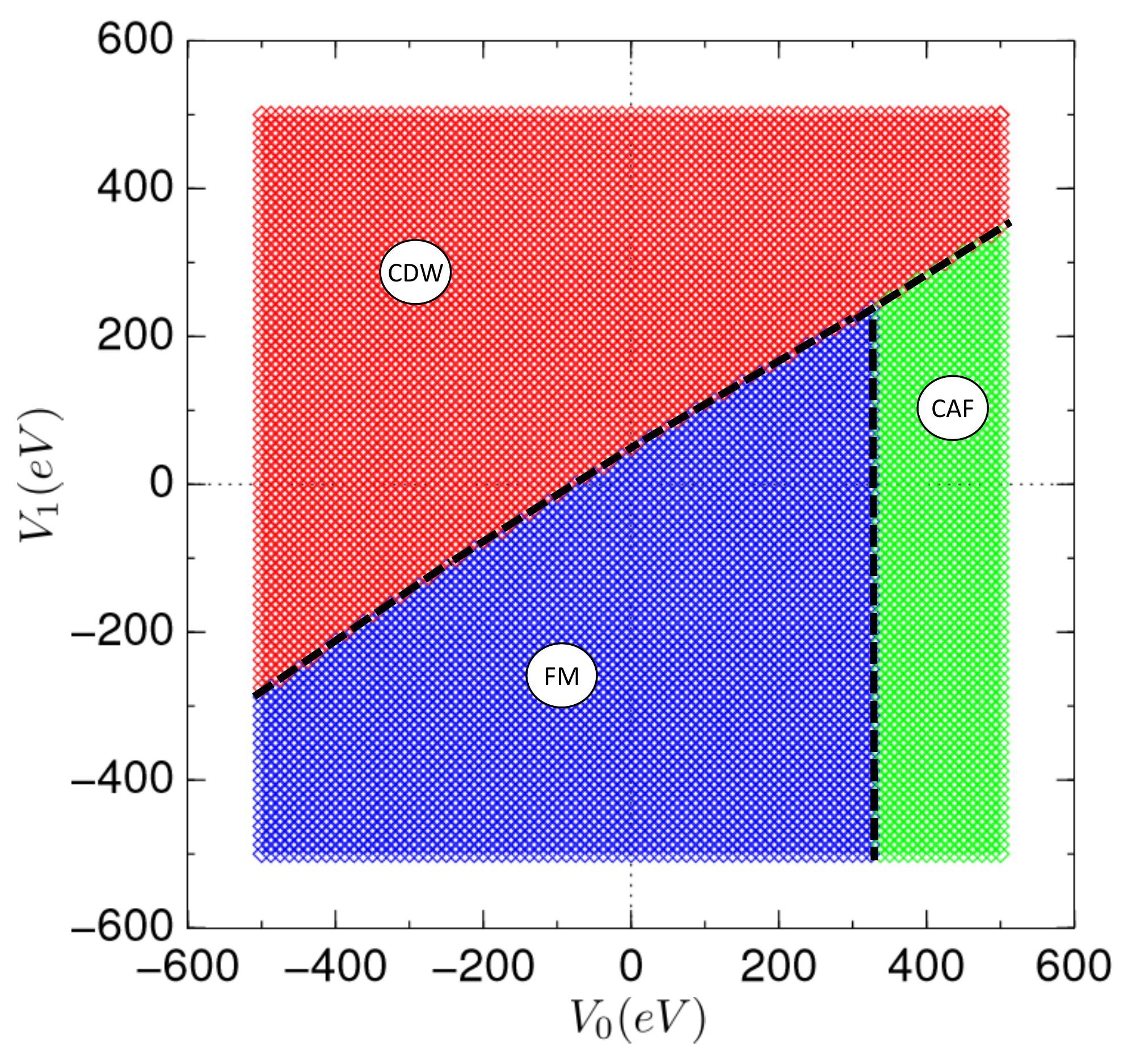}
    \caption{}
    \label{Figure5a}
  \end{subfigure}
  \hspace{20pt}
  \begin{subfigure}[b]{0.4\textwidth}
    \includegraphics[width=\textwidth]{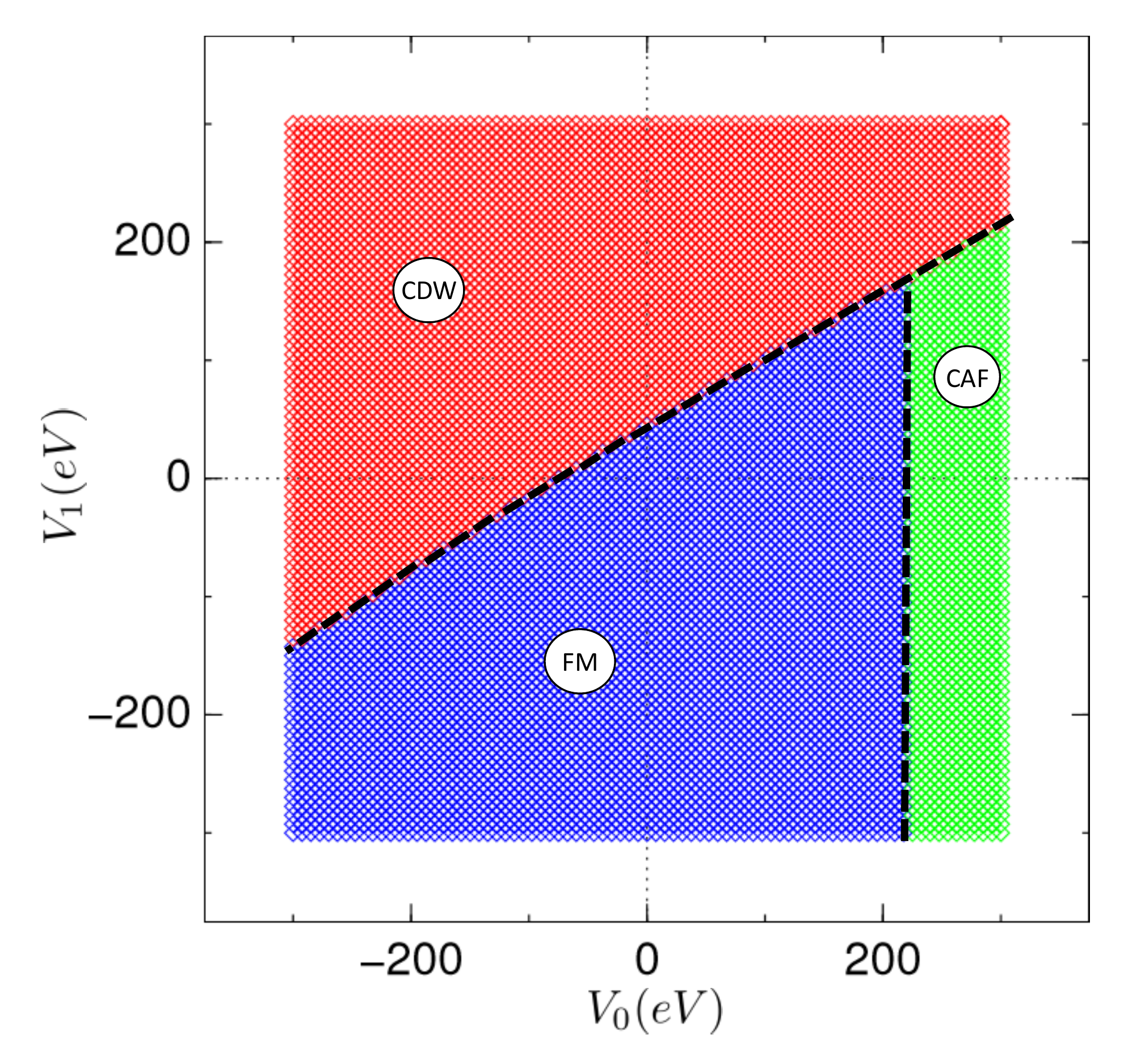}
   \caption{}
    \label{Figure5b}
  \end{subfigure}
  \caption{(a) Phase diagram for $N=3$ with $B_T=30T$. The boundary separating the CAF (green) and FM (blue) phases is governed by a continuous phase transition. The CDW (red) remains in a large region of parameter space where the on-site interaction is attractive. (b) Phase diagram for $N=5$ again for $B_T=30T$. Notice that the qualitative behavior between these two diagrams remains quite similar, however the CAF-FM phase boundary in the $N=5$ diagram has shifted down by almost 100eV (along the $V_0$-axis) from where it is for $N=3$.}
\end{figure}

The inclusion of more LLs in the active window yields a much more interesting phase diagram. For $N=3$, the three active LLs are $n=-1,0,1$. This is also the smallest possible window size which is impacted in a non-trivial way by the filled sea. Figure \ref{Figure4} shows the $N=3$ phase diagram with vanishing Zeeman coupling ($\xi_z=0$), in which the AFM (green) now appears. The phase boundaries in Fig. \ref{Figure4} are all first order transitions. When a non-zero Zeeman field is included ($B_T=30T$), shown in Fig. \ref{Figure5a}, the AFM phase is replaced by the CAF (green). Again, the FM state takes up more space in the phase diagram. The phase boundary separating the FM and CAF states now corresponds to a continuous phase transition, in which the canting evolves into a full spin polarization along the direction of the total field. The increasing parameter space of the FM state with increasing Zeeman coupling is consistent with recent transport experiments \cite{Young2014}.

Assuming that the real microscopic interaction parameters for graphene are both positive, the physically relevant region should be somewhere in the first quadrant of this phase diagram. Note that at this level of approximation (small window size), the CAF-FM phase boundary occurs at unphysically large values of $V_0$. This issue is resolved by including more LLs in the active window. For example, $V_0-V_1$ phase diagrams for window sizes of $N=5, 7, 9$, and $11$ were also computed. In each case, we saw roughly the same qualitative picture, with the exception that the CAF-FM phase boundary moves to smaller values of $V_0$, with increasing window size. Figure \ref{Figure5a}-b shows a comparison of this behavior for $N=3$ and $N=5$. Notice that the value of $V_0$ at the phase boundary is roughly 100eV less than in the $N=3$ case. This rather large quantitative change in the phase boundary suggests that, ideally, one should include all the LL's in the active window. However, for $B_\perp=15T$, the required window size would be $N=5261$ ($n_c=n_w=2630$). Including this many LL's is computationally prohibitive, so an extrapolation method was used, as discussed in Sec. \ref{extrapolation} below.
\subsection{Stability of the AFM/CAF state}
It is interesting that the CAF state is only stable for $N \geq 3$, indicating that LLM is crucial to its presence in the phase diagram. This can be understood in further detail for the $N=3$ case. For simplicity, we consider a model with only OS interaction energy, taking $V_1=\xi_z=0$. Under these conditions, the AFM and FM states are both degnerate when the OS interaction is neglected. For the HF state, noting that there is no admixing of spin or valley indices in the density matrix for the AFM and FM states, one finds
\beq
\braket{\mathcal{H}^{OS}}=\frac{V_0a_c}{4}\int d^2R\sum_{s}n_{\uparrow}(\bo{r},s)n_{\downarrow}(\bo{r},s),
\eeq
where
\beq
n_{\sigma}(\bo{r},s)\equiv\sum_{\tau}\braket{\Psi^{\dagger}_{\sigma,\tau}(\bo{r},s)\Psi_{\sigma,\tau}(\bo{r},s)}
\eeq
is the spin-resolved density. Expanding the fields in terms of the LL eigenstates (Eq. \ref{2.1.5}) and separating terms which are diagonal from those that are off diagonal, one finds
\beq
\begin{aligned}
&n_{\sigma}(s=A)=\frac{1}{2(2\pi\ell^2)}\bigg\{\sum_{n=-1}^1\big[2^{\delta_{n,0}}\rho(n,K,\sigma;n,K,\sigma)+(1-\delta_{n,0}) \ \rho(n,K^{\prime},\sigma;n,K^{\prime},\sigma) \big]\\
&+\frac{1}{2}\big[\rho(-1,K,\sigma;1,K,\sigma)-\rho(-1,K^{\prime},\sigma;1,K^{\prime},\sigma)+\rho(1,K,\sigma;-1,K,\sigma)-\rho(1,K^{\prime},\sigma;-1,K^{\prime},\sigma) \big]\bigg\}.\ \label{occupationdensityA}
\end{aligned}
\eeq
Interchanging $K^{\prime} \leftrightarrow K$ in $n_{\sigma}(s=A)$ yields $n_{\sigma}(s=B)$. In the $N=3$ case, $\rho(-1,\tau,\sigma;-1,\tau,\sigma)+\rho(1,\tau,\sigma;1,\tau,\sigma)=1$ for both the AFM and FM states. There is also a small, but non-zero, amount of LLM within the density matrix, and these off-diagonal (in LL index) elements have a particular sign signature which differs between the two states:
\beq
\begin{aligned}
&\rho^{FM}(1,K,\uparrow;-1,K,\uparrow)=a_1 \hspace{16pt},\hspace{10pt} \rho^{AFM}(1,K,\uparrow;-1,K,\uparrow)=a_2;\\
&\rho^{FM}(1,K,\downarrow;-1,K,\downarrow)=-b_1 \hspace{10pt},\hspace{10pt} \rho^{AFM}(1,K,\downarrow;-1,K,\downarrow)=-b_2;\\
&\rho^{FM}(1,K^{\prime},\uparrow;-1,K^{\prime},\uparrow)=a_1 \hspace{10pt},\hspace{10pt} \rho^{AFM}(1,K^{\prime},\uparrow;-1,K^{\prime},\uparrow)=-b_2;\\
&\rho^{FM}(1,K^{\prime},\downarrow;-1,K^{\prime},\downarrow)=-b_1 \hspace{4pt},\hspace{10pt} \rho^{AFM}(1,K^{\prime},\downarrow;-1,K^{\prime},\downarrow)=a_2. \ \label{signsignature}
\end{aligned}
\eeq
In Eq. \ref{signsignature}, $a_1$, $a_2$, $b_1$, and $b_2$ are positive constants, and are the results of minimizing the energy for a HF state. As a direct consequence of the sign signatures in Eq. \ref{signsignature}, the LLM terms for the FM spin-resolved density, on both sublattices, cancel, yielding
\beq
\begin{aligned}
n&^{FM}_{\sigma}(A)=\frac{1}{2\pi\ell^2}\bigg\{\rho^{FM}(0,K,\sigma;0,K,\sigma)+1\bigg\},\\
n&^{FM}_{\sigma}(B)=\frac{1}{2\pi\ell^2}\bigg\{\rho^{FM}(0,K^{\prime},\sigma;0,K^{\prime},\sigma)+1\bigg\}.
\end{aligned}
\eeq
Using Eq. \ref{3.1.1} for the zLL structure of the FM, the OS energy is 
\beq
\begin{aligned}
\braket{\mathcal{H}^{OS}}^{FM}=\frac{V_0a_c}{(2\pi\ell^2)}\frac{L_xL_y}{2\pi\ell^2}=g\frac{V_0a_c}{2\pi\ell^2},
\end{aligned}
\eeq
where $g=\frac{L_xL_y}{2\pi\ell^2}$ is the degeneracy per LL. 

For the AFM state, LLM has a non-trivial effect on the spin-resolved density:
\beq
\begin{aligned}
n_{\uparrow}^{AFM}(A)&=\frac{1}{2\pi\ell^2}\big[2+(a_2+b_2)\big] \ , \ n_{\downarrow}^{AFM}(A)=\frac{1}{2\pi\ell^2}\big[1-(a_2+b_2)\big];\\
n_{\uparrow}^{AFM}(B)&=\frac{1}{2\pi\ell^2}\big[1-(a_2+b_2)\big] \ , \ n_{\downarrow}^{AFM}(B)=\frac{1}{2\pi\ell^2}\big[2+(a_2+b_2)\big].
\end{aligned}
\eeq
The OS interaction energy for the AFM state is then
\beq
\begin{aligned}
\braket{\mathcal{H}^{OS}}^{AFM}=g\frac{V_0a_c}{2\pi\ell^2}\bigg[1-\frac{a_2+b_2}{2}-\frac{(a_2+b_2)^2}{2} \bigg] < g\frac{V_0a_c}{2\pi\ell^2}=\braket{\mathcal{H}^{OS}}^{FM},
\end{aligned}
\eeq
for positive $V_0$. 

\begin{figure}[h]
\centering
  \begin{subfigure}[b]{0.35\textwidth}
    \includegraphics[width=\textwidth]{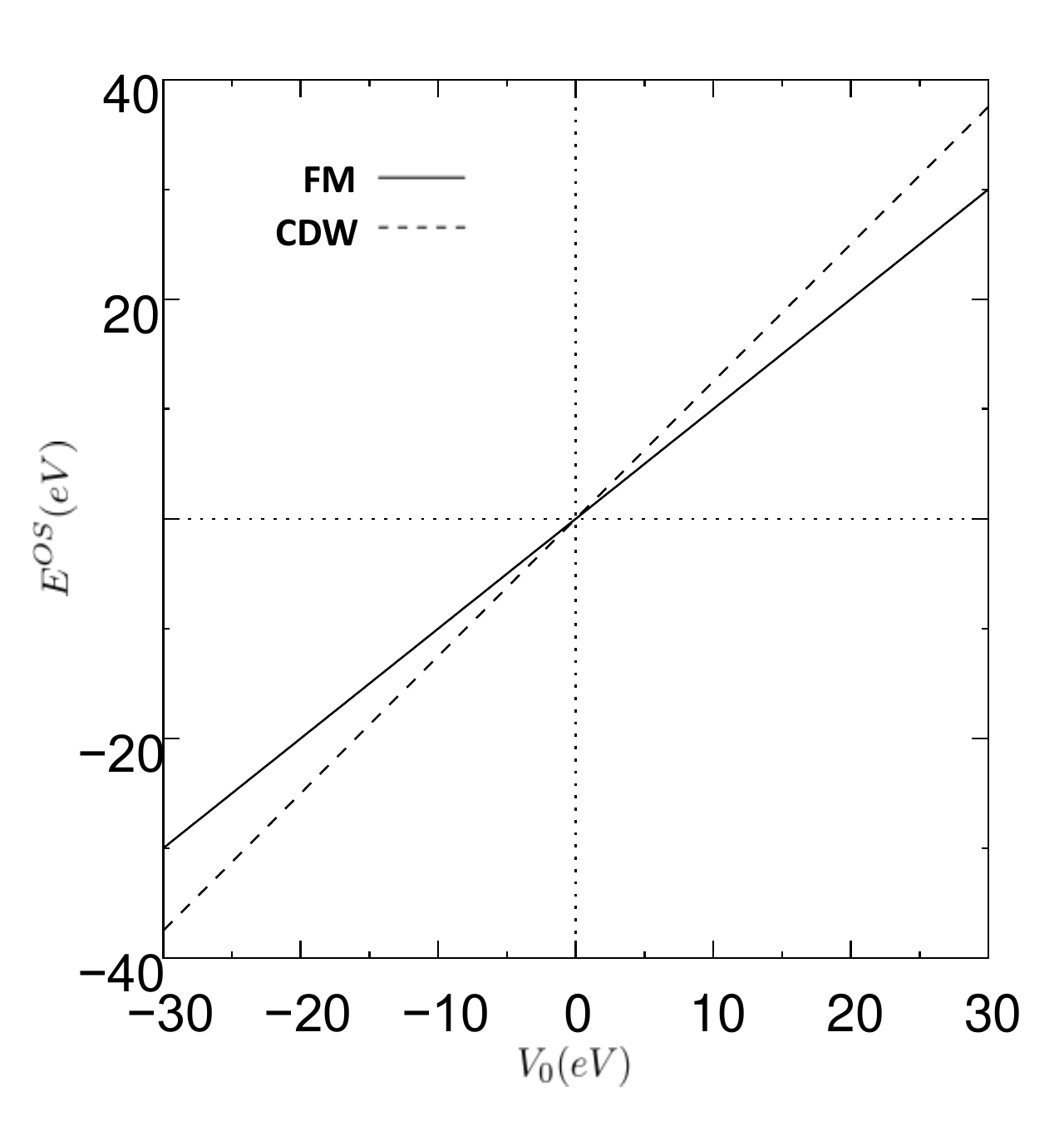}
    \caption{}
    \label{Figure6a}
  \end{subfigure}
  \hspace{20pt}
  \begin{subfigure}[b]{0.35\textwidth}
    \includegraphics[width=\textwidth]{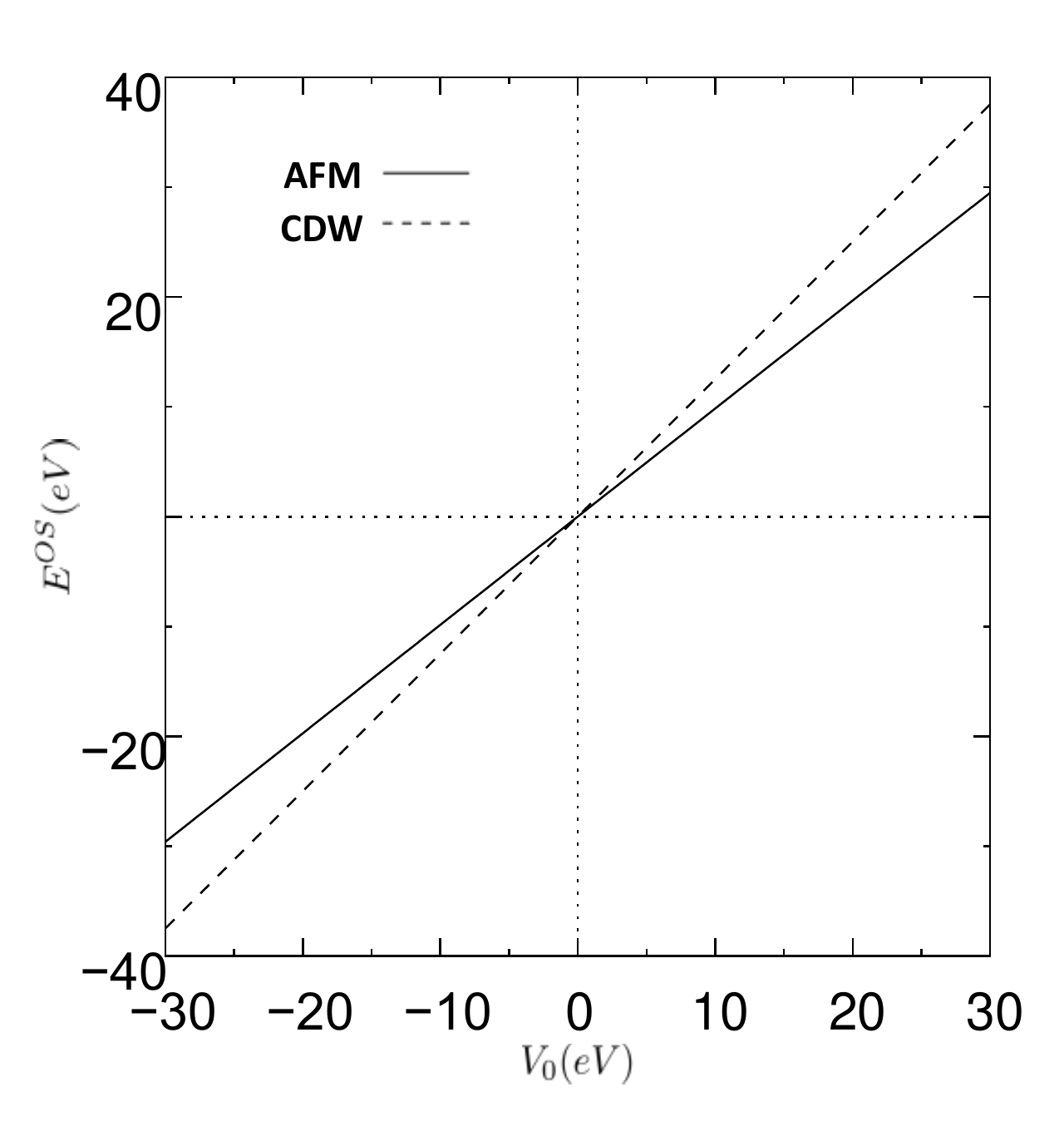}
   \caption{}
    \label{Figure6b}
  \end{subfigure}\\
  \begin{subfigure}[b]{0.35\textwidth}
    \includegraphics[width=\textwidth]{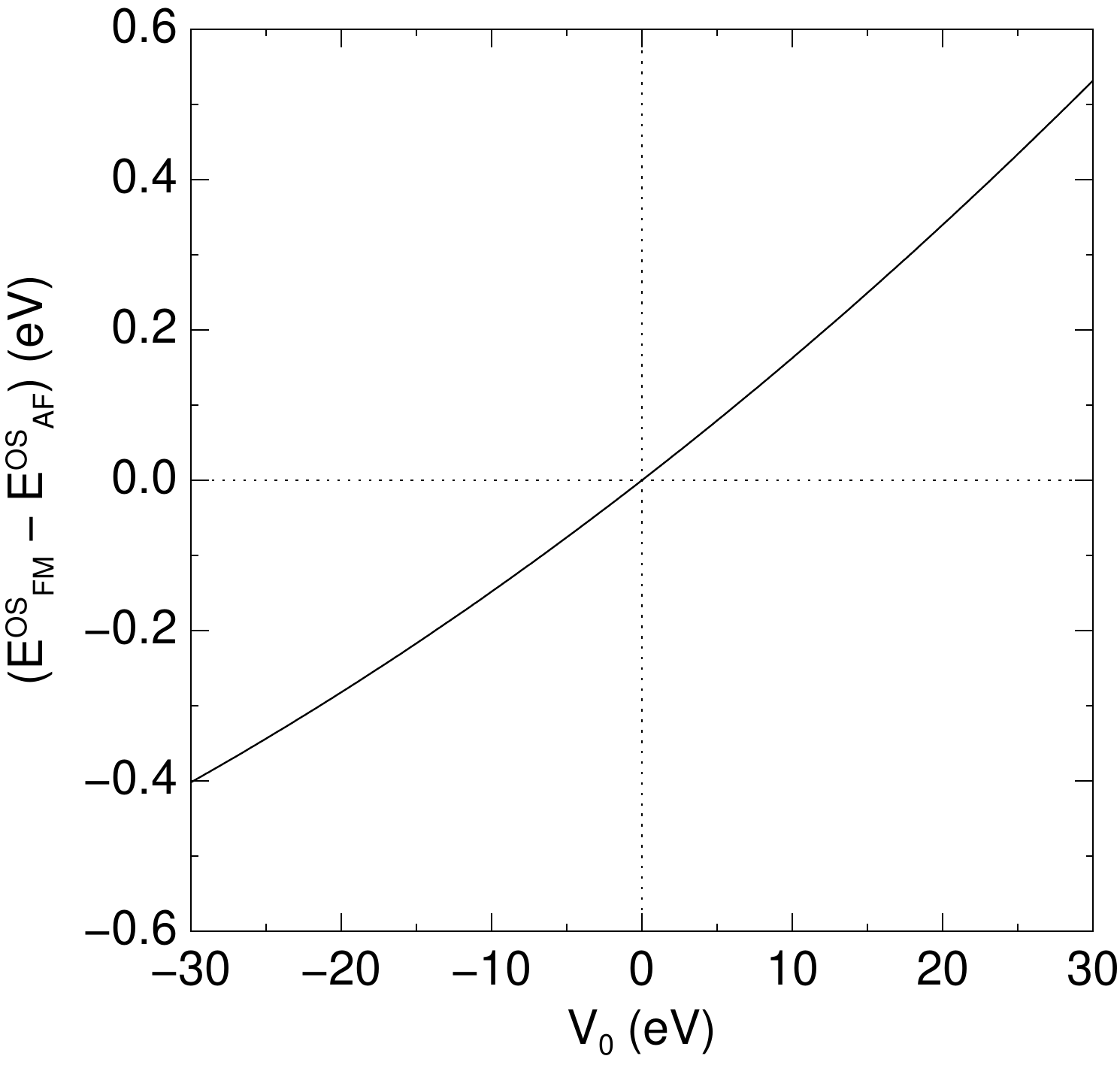}
   \caption{}
    \label{Figure6c}
  \end{subfigure}
  \caption{(a,b) Comparison of on-site energy, per guiding center index ($E^{OS}_i=\frac{2\pi\ell^2}{a_cg}\braket{\mathcal{H}^{OS}}^i$), of the CDW-FM and CDW-AFM states. For $V_0<0$, the CDW has the lowest of the three energies. When $V_0 > 0$, the AFM becomes the favored state. Note that the FM and AFM energies are very close over this range of $V_0$. (c) On-site energy difference between the FM and AFM states. The energy of the AFM is clearly less than the FM for $V_0 > 0$. This energy gap becomes greater for larger $V_0$ and/or larger window sizes ($N>3$).}
\end{figure}

This example shows that the LLM contributions for the FM vanish completely because of the LLM sign signature in its density matrix. Analogous behavior occurs in the sign signatures for the LLM in the FM state for larger values of N. In essence this means the FM state cannot take advantage of LLM to lower its energy, whereas the AFM has more flexibility. For this reason, $\braket{\mathcal{H^{OS}}}^{AFM} < \braket{\mathcal{H}^{OS}}^{FM}$. While the CAF state is more complex than the AFM state, it similarly can take advantage of LLM so that, for small $\xi_z$, the CAF has lower energy than the FM. With increasing $\xi_z$, this energetic advantage is ultimately lost and the FM becomes the groundstate. Figure \ref{Figure6a}-c summarizes these observations.
\subsection{$\xi_z - V_0$ phase diagrams}
\begin{figure}[h]
\centering
\captionbox
{Phase diagram for $N=7$ and $B_{\perp}=15T$ within the space of the Zeeman field and on-site interaction strengths. The three phases here correspond to the CDW (red), FM (blue), and CAF (green).\label{Figure7}}
{\includegraphics[width=0.55\linewidth]{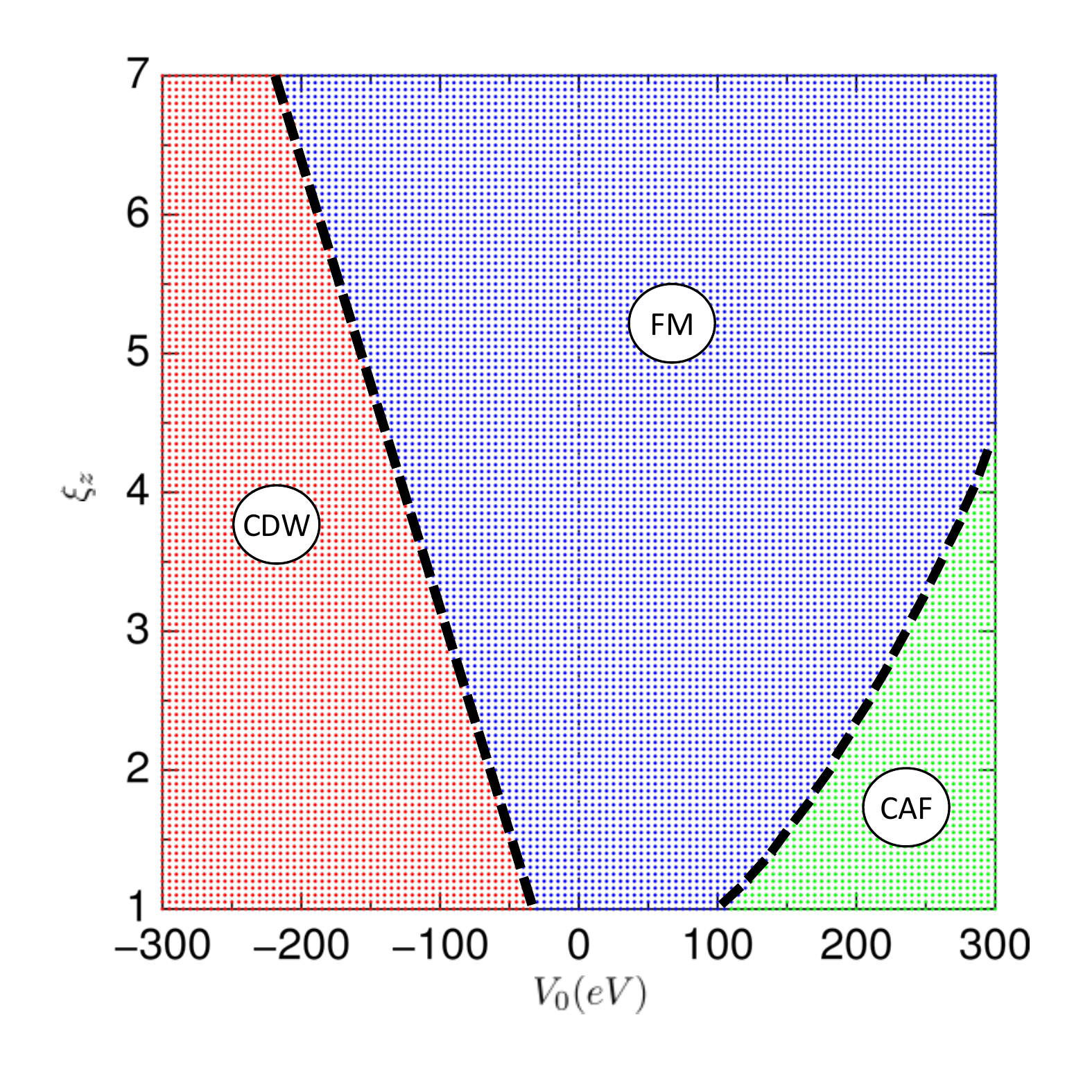}}
\end{figure}

\begin{figure}[h]
\centering
\captionbox
{Phase boundaries for window sizes $N=57$, $119$, and $301$. The renormalization of the boundary is quite extreme as $N$ is increased.\label{Figure8}}
{\includegraphics[width=0.5\linewidth]{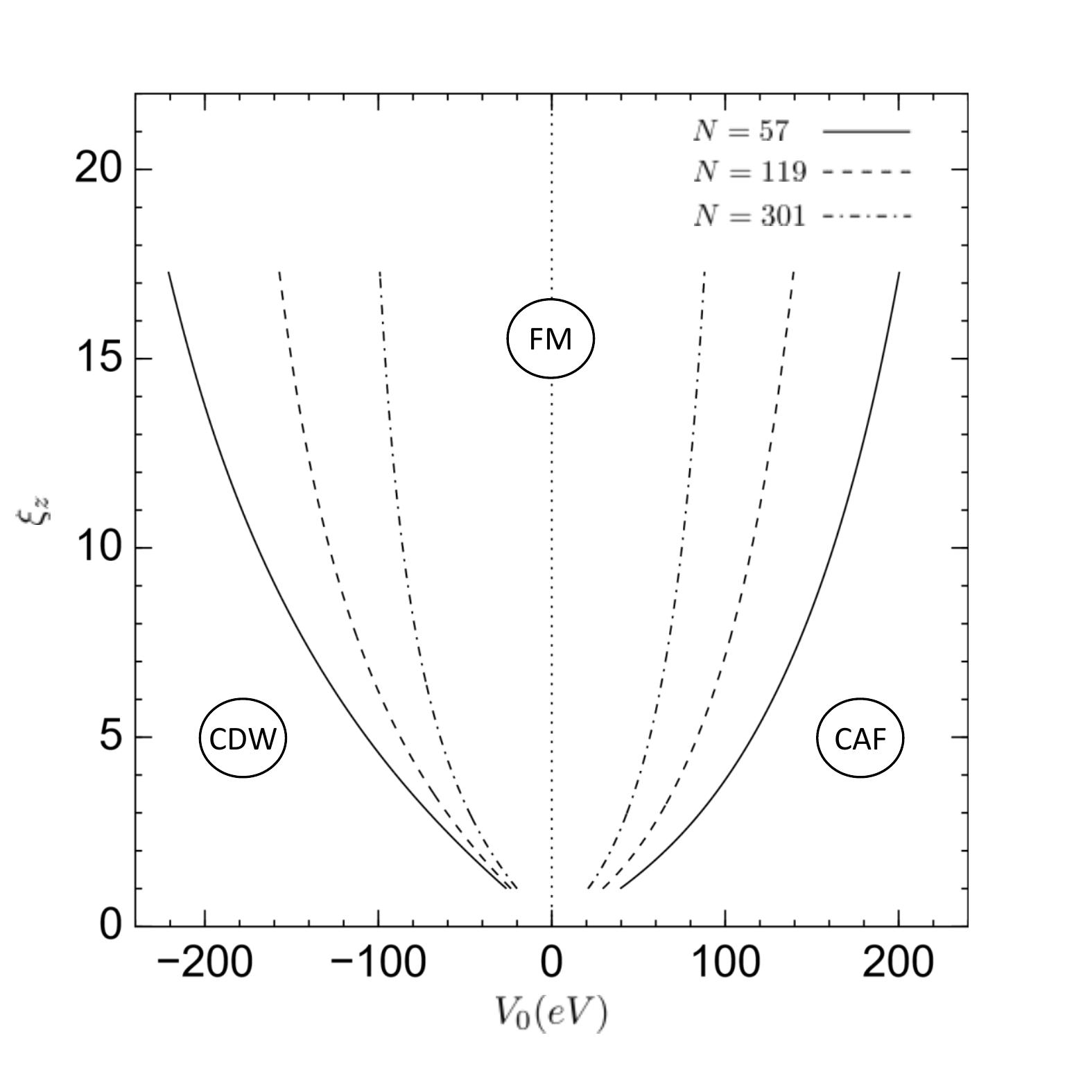}}
\end{figure}

Phase diagrams were also constructed within the space of $\xi_z$ and $V_0$, for $V_1=0$. Figure \ref{Figure7} shows such a phase diagram for $N=7$. The CAF-FM phase boundary occurs again around unphysically large values of $V_0$.
Figure \ref{Figure8} shows the phase boundaries for much larger window sizes. Note the boundaries move significantly as $N$ is increased. Within the context of this model, calculating the precise phase boundary location would require a window size of $N=5261$. However, we were not able to exceed $N=301$ due to computational constraints. To estimate the positions of the physical phase boundaries, we extrapolated our results, for computationally accessible values of $N$, out to this larger value.
\subsubsection{Extrapolation} \label{extrapolation}
\begin{figure}[h]
\centering
\captionbox
{Phase diagram formed by extrapolating to a window size of $N=5261$. \label{Figure9}}
{\includegraphics[width=0.55\linewidth]{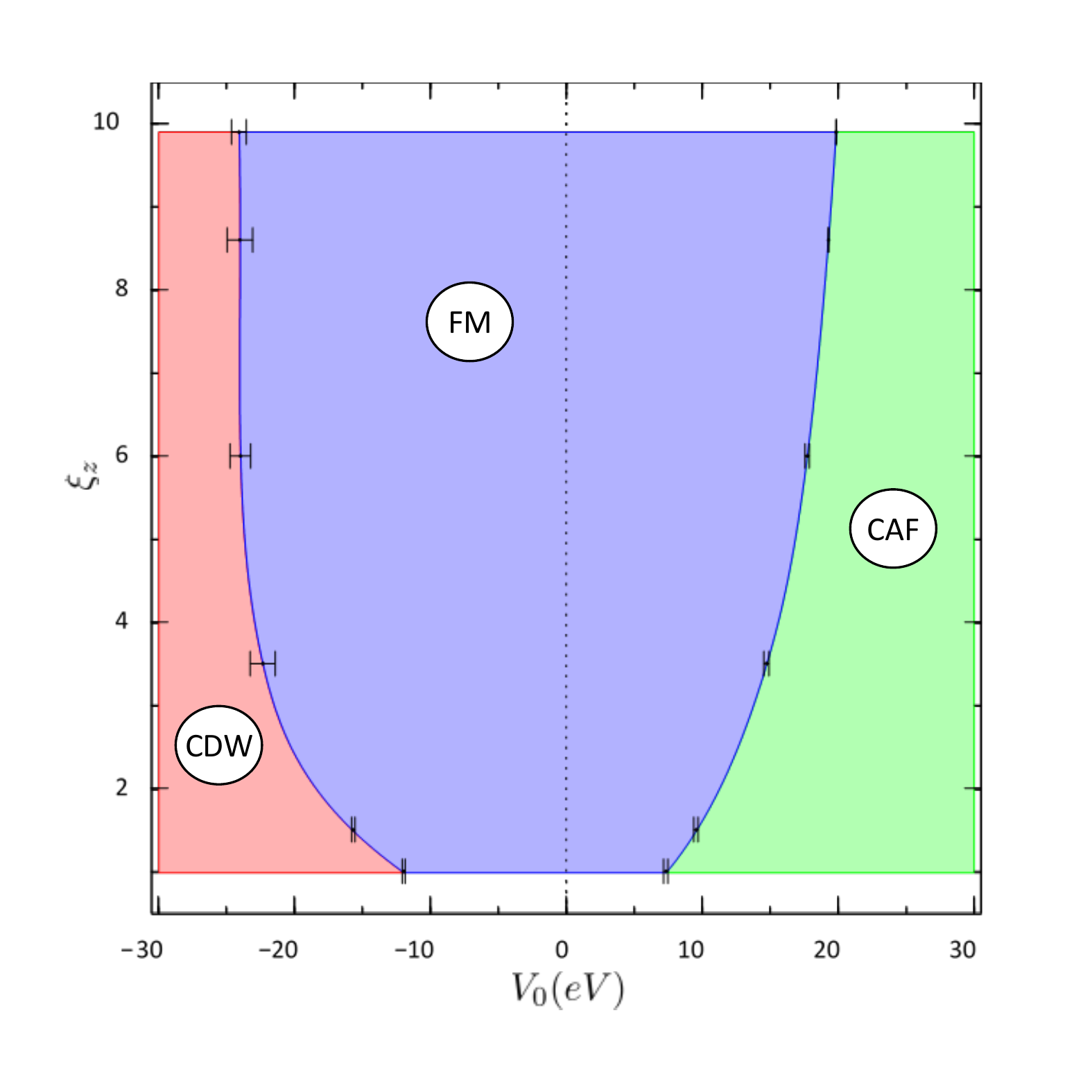}}
\end{figure}

The $V_0$ transition values along both phase boundaries were collected for several window sizes ($57\leq N \leq 301$) over a range of Zeeman fields ($1.0\leq \xi_z \leq17.3$). The data was then fitted to a power law, $V_0(N)=a/N^b$, at fixed Zeeman field, for the CDW-FM and CAF-FM phase boundaries, where $a$ and $b$ are fitting parameters. The fit for the CAF-FM phase boundary followed this power law behavior quite well. For the CDW-FM boundary, the power law behavior only emerged at larger window sizes, increasing the uncertainty in the extrapolation. Figure \ref{Figure9} shows the results for both phase boundaries. The error bars were determined by using different sample sizes of data points in the extrapolation.
The physically important insight from this window size study is that, while small window sizes capture the correct qualitative behavior of the phase boundaries (except for $N=1$), by themselves, they yield poor estimates of the critical parameters, greatly overestimating the energy scales at which transitions occur. This is dramatically illustrated by compairing the $N=7$ results (Fig. \ref{Figure7}) with our extrapolated results (Fig. \ref{Figure9}), for which the critical values of $V_0$ differ by an order of magnitude.

Using the estimate found in Ref. \onlinecite{Katsnelson2011} for the OS Hubbard interaction strength ($V_0=9.3eV$), we note that the CAF-FM phase boundary passes right through a range of accessible Zeeman fields. This supports the interpretation of the conductance transition as a function of Zeeman field, observed in Ref. \onlinecite{Young2014}, in terms of a CAF-FM transition.
\section{Conclusion}
In summary, we have presented results of numerical Hartree-Fock calculations for a graphene model of the $\nu=0$ quantum Hall state, with both short-range lattice and long-range Coulomb interactions. This model supports three phases (FM, CDW, and CAF) in the $V_0-V_1-\xi_z$ parameter space for $N \geq 3$. For non-zero Zeeman energy, when the filled sea is not included ($N=1$), only the FM and the CDW are stable, minimum-energy, solutions. As more LLs are included in the active window, the phase boundaries become strongly renormalized, moving into a physically relevant region of the phase diagram.

LLM plays a substantial role in determining which phase is the groundstate. This is ultimately determined by the sign signature which appears in the off-diagonal elements of the density matrix for a particular state, allowing states with AF order to take advantage of LLM in a way that the FM cannot.
\section{Acknowledgements}
The authors are grateful to Ganpathy Murthy, Efrat Shimshoni, and Andrea Young for helpful conversations. This work was supported by the NSF through Grant Nos. DMR-1506263 and DMR-1506460, and by the US-Israel Binational Science Foundation through Grant No. 2012120. HAF also thanks the Aspen Center for Physics (NSF Grant No. 1066293) for its hospitality.

\providecommand{\noopsort}[1]{}\providecommand{\singleletter}[1]{#1}%
%


\begin{thebibliography}{24}%
\makeatletter
\providecommand \@ifxundefined [1]{%
 \@ifx{#1\undefined}
}%
\providecommand \@ifnum [1]{%
 \ifnum #1\expandafter \@firstoftwo
 \else \expandafter \@secondoftwo
 \fi
}%
\providecommand \@ifx [1]{%
 \ifx #1\expandafter \@firstoftwo
 \else \expandafter \@secondoftwo
 \fi
}%
\providecommand \natexlab [1]{#1}%
\providecommand \enquote  [1]{``#1''}%
\providecommand \bibnamefont  [1]{#1}%
\providecommand \bibfnamefont [1]{#1}%
\providecommand \citenamefont [1]{#1}%
\providecommand \href@noop [0]{\@secondoftwo}%
\providecommand \href [0]{\begingroup \@sanitize@url \@href}%
\providecommand \@href[1]{\@@startlink{#1}\@@href}%
\providecommand \@@href[1]{\endgroup#1\@@endlink}%
\providecommand \@sanitize@url [0]{\catcode `\\12\catcode `\$12\catcode
  `\&12\catcode `\#12\catcode `\^12\catcode `\_12\catcode `\%12\relax}%
\providecommand \@@startlink[1]{}%
\providecommand \@@endlink[0]{}%
\providecommand \url  [0]{\begingroup\@sanitize@url \@url }%
\providecommand \@url [1]{\endgroup\@href {#1}{\urlprefix }}%
\providecommand \urlprefix  [0]{URL }%
\providecommand \Eprint [0]{\href }%
\providecommand \doibase [0]{http://dx.doi.org/}%
\providecommand \selectlanguage [0]{\@gobble}%
\providecommand \bibinfo  [0]{\@secondoftwo}%
\providecommand \bibfield  [0]{\@secondoftwo}%
\providecommand \translation [1]{[#1]}%
\providecommand \BibitemOpen [0]{}%
\providecommand \bibitemStop [0]{}%
\providecommand \bibitemNoStop [0]{.\EOS\space}%
\providecommand \EOS [0]{\spacefactor3000\relax}%
\providecommand \BibitemShut  [1]{\csname bibitem#1\endcsname}%
\let\auto@bib@innerbib\@empty
\bibitem [{\citenamefont {Novoselov}\ \emph {et~al.}(2004)\citenamefont
  {Novoselov}, \citenamefont {Geim}, \citenamefont {Morozov}, \citenamefont
  {Jiang}, \citenamefont {Zhang}, \citenamefont {Dubonos}, \citenamefont
  {Grigorieva},\ and\ \citenamefont {Firsov}}]{Novoselov2004}%
  \BibitemOpen
  \bibfield  {author} {\bibinfo {author} {\bibfnamefont {K.~S.}\ \bibnamefont
  {Novoselov}}, \bibinfo {author} {\bibfnamefont {A.~K.}\ \bibnamefont {Geim}},
  \bibinfo {author} {\bibfnamefont {S.~V.}\ \bibnamefont {Morozov}}, \bibinfo
  {author} {\bibfnamefont {D.}~\bibnamefont {Jiang}}, \bibinfo {author}
  {\bibfnamefont {Y.}~\bibnamefont {Zhang}}, \bibinfo {author} {\bibfnamefont
  {S.~V.}\ \bibnamefont {Dubonos}}, \bibinfo {author} {\bibfnamefont {I.~V.}\
  \bibnamefont {Grigorieva}}, \ and\ \bibinfo {author} {\bibfnamefont {A.~A.}\
  \bibnamefont {Firsov}},\ }\href {\doibase 10.1126/science.1102896} {\bibfield
   {journal} {\bibinfo  {journal} {Science}\ }\textbf {\bibinfo {volume}
  {306}},\ \bibinfo {pages} {666} (\bibinfo {year} {2004})},\ \Eprint
  {http://arxiv.org/abs/http://science.sciencemag.org/content/306/5696/666.full.pdf}
  {http://science.sciencemag.org/content/306/5696/666.full.pdf} \BibitemShut
  {NoStop}%
\bibitem [{\citenamefont {Zhang}\ \emph {et~al.}(2005)\citenamefont {Zhang},
  \citenamefont {Tan}, \citenamefont {Stormer},\ and\ \citenamefont
  {Kim}}]{Kim2005}%
  \BibitemOpen
  \bibfield  {author} {\bibinfo {author} {\bibfnamefont {Y.}~\bibnamefont
  {Zhang}}, \bibinfo {author} {\bibfnamefont {Y.-W.}\ \bibnamefont {Tan}},
  \bibinfo {author} {\bibfnamefont {H.~L.}\ \bibnamefont {Stormer}}, \ and\
  \bibinfo {author} {\bibfnamefont {P.}~\bibnamefont {Kim}},\ }\href@noop {}
  {\bibfield  {journal} {\bibinfo  {journal} {Nature}\ }\textbf {\bibinfo
  {volume} {438}},\ \bibinfo {pages} {201} (\bibinfo {year}
  {2005})}\BibitemShut {NoStop}%
\bibitem [{\citenamefont {Novoselov}\ \emph {et~al.}(2005)\citenamefont
  {Novoselov}, \citenamefont {Geim}, \citenamefont {Morozov}, \citenamefont
  {Jiang}, \citenamefont {Katsnelson}, \citenamefont {Grigorieva},
  \citenamefont {Dubonos},\ and\ \citenamefont {Firsov}}]{Novoselov2005}%
  \BibitemOpen
  \bibfield  {author} {\bibinfo {author} {\bibfnamefont {K.~S.}\ \bibnamefont
  {Novoselov}}, \bibinfo {author} {\bibfnamefont {A.~K.}\ \bibnamefont {Geim}},
  \bibinfo {author} {\bibfnamefont {S.~V.}\ \bibnamefont {Morozov}}, \bibinfo
  {author} {\bibfnamefont {D.}~\bibnamefont {Jiang}}, \bibinfo {author}
  {\bibfnamefont {M.~I.}\ \bibnamefont {Katsnelson}}, \bibinfo {author}
  {\bibfnamefont {I.~V.}\ \bibnamefont {Grigorieva}}, \bibinfo {author}
  {\bibfnamefont {S.~V.}\ \bibnamefont {Dubonos}}, \ and\ \bibinfo {author}
  {\bibfnamefont {A.~A.}\ \bibnamefont {Firsov}},\ }\href@noop {} {\bibfield
  {journal} {\bibinfo  {journal} {Nature}\ }\textbf {\bibinfo {volume} {438}},\
  \bibinfo {pages} {197} (\bibinfo {year} {2005})}\BibitemShut {NoStop}%
\bibitem [{\citenamefont {Jiang}\ \emph {et~al.}(2007)\citenamefont {Jiang},
  \citenamefont {Zhang}, \citenamefont {Stormer},\ and\ \citenamefont
  {Kim}}]{Kim2007}%
  \BibitemOpen
  \bibfield  {author} {\bibinfo {author} {\bibfnamefont {Z.}~\bibnamefont
  {Jiang}}, \bibinfo {author} {\bibfnamefont {Y.}~\bibnamefont {Zhang}},
  \bibinfo {author} {\bibfnamefont {H.~L.}\ \bibnamefont {Stormer}}, \ and\
  \bibinfo {author} {\bibfnamefont {P.}~\bibnamefont {Kim}},\ }\href {\doibase
  10.1103/PhysRevLett.99.106802} {\bibfield  {journal} {\bibinfo  {journal}
  {Phys. Rev. Lett.}\ }\textbf {\bibinfo {volume} {99}},\ \bibinfo {pages}
  {106802} (\bibinfo {year} {2007})}\BibitemShut {NoStop}%
\bibitem [{\citenamefont {Young}\ \emph {et~al.}(2012)\citenamefont {Young},
  \citenamefont {Dean}, \citenamefont {Wang}, \citenamefont {Ren},
  \citenamefont {Cadden-Zimansky}, \citenamefont {Watanabe}, \citenamefont
  {Taniguchi}, \citenamefont {Hone}, \citenamefont {Shepard},\ and\
  \citenamefont {Kim}}]{Young2012}%
  \BibitemOpen
  \bibfield  {author} {\bibinfo {author} {\bibfnamefont {A.~F.}\ \bibnamefont
  {Young}}, \bibinfo {author} {\bibfnamefont {C.~R.}\ \bibnamefont {Dean}},
  \bibinfo {author} {\bibfnamefont {L.}~\bibnamefont {Wang}}, \bibinfo {author}
  {\bibfnamefont {H.}~\bibnamefont {Ren}}, \bibinfo {author} {\bibfnamefont
  {P.}~\bibnamefont {Cadden-Zimansky}}, \bibinfo {author} {\bibfnamefont
  {K.}~\bibnamefont {Watanabe}}, \bibinfo {author} {\bibfnamefont
  {T.}~\bibnamefont {Taniguchi}}, \bibinfo {author} {\bibfnamefont
  {J.}~\bibnamefont {Hone}}, \bibinfo {author} {\bibfnamefont {K.~L.}\
  \bibnamefont {Shepard}}, \ and\ \bibinfo {author} {\bibfnamefont
  {P.}~\bibnamefont {Kim}},\ }\href@noop {} {\bibfield  {journal} {\bibinfo
  {journal} {Nature Physics}\ }\textbf {\bibinfo {volume} {8}},\ \bibinfo
  {pages} {550} (\bibinfo {year} {2012})}\BibitemShut {NoStop}%
\bibitem [{\citenamefont {Yang}\ \emph {et~al.}(2006)\citenamefont {Yang},
  \citenamefont {Das~Sarma},\ and\ \citenamefont {MacDonald}}]{DasSarma2006}%
  \BibitemOpen
  \bibfield  {author} {\bibinfo {author} {\bibfnamefont {K.}~\bibnamefont
  {Yang}}, \bibinfo {author} {\bibfnamefont {S.}~\bibnamefont {Das~Sarma}}, \
  and\ \bibinfo {author} {\bibfnamefont {A.~H.}\ \bibnamefont {MacDonald}},\
  }\href {\doibase 10.1103/PhysRevB.74.075423} {\bibfield  {journal} {\bibinfo
  {journal} {Phys. Rev. B}\ }\textbf {\bibinfo {volume} {74}},\ \bibinfo
  {pages} {075423} (\bibinfo {year} {2006})}\BibitemShut {NoStop}%
\bibitem [{\citenamefont {Basko}\ and\ \citenamefont
  {Aleiner}(2008)}]{Basko2008}%
  \BibitemOpen
  \bibfield  {author} {\bibinfo {author} {\bibfnamefont {D.~M.}\ \bibnamefont
  {Basko}}\ and\ \bibinfo {author} {\bibfnamefont {I.~L.}\ \bibnamefont
  {Aleiner}},\ }\href {\doibase 10.1103/PhysRevB.77.041409} {\bibfield
  {journal} {\bibinfo  {journal} {Phys. Rev. B}\ }\textbf {\bibinfo {volume}
  {77}},\ \bibinfo {pages} {041409} (\bibinfo {year} {2008})}\BibitemShut
  {NoStop}%
\bibitem [{\citenamefont {Kharitonov}(2012)}]{Kharitonov2012}%
  \BibitemOpen
  \bibfield  {author} {\bibinfo {author} {\bibfnamefont {M.}~\bibnamefont
  {Kharitonov}},\ }\href {\doibase 10.1103/PhysRevB.85.155439} {\bibfield
  {journal} {\bibinfo  {journal} {Phys. Rev. B}\ }\textbf {\bibinfo {volume}
  {85}},\ \bibinfo {pages} {155439} (\bibinfo {year} {2012})}\BibitemShut
  {NoStop}%
\bibitem [{\citenamefont {Abanin}\ \emph {et~al.}(2006)\citenamefont {Abanin},
  \citenamefont {Lee},\ and\ \citenamefont {Levitov}}]{Levitov2006}%
  \BibitemOpen
  \bibfield  {author} {\bibinfo {author} {\bibfnamefont {D.~A.}\ \bibnamefont
  {Abanin}}, \bibinfo {author} {\bibfnamefont {P.~A.}\ \bibnamefont {Lee}}, \
  and\ \bibinfo {author} {\bibfnamefont {L.~S.}\ \bibnamefont {Levitov}},\
  }\href {\doibase 10.1103/PhysRevLett.96.176803} {\bibfield  {journal}
  {\bibinfo  {journal} {Phys. Rev. Lett.}\ }\textbf {\bibinfo {volume} {96}},\
  \bibinfo {pages} {176803} (\bibinfo {year} {2006})}\BibitemShut {NoStop}%
\bibitem [{\citenamefont {Fertig}\ and\ \citenamefont
  {Brey}(2006)}]{Fertig2006}%
  \BibitemOpen
  \bibfield  {author} {\bibinfo {author} {\bibfnamefont {H.~A.}\ \bibnamefont
  {Fertig}}\ and\ \bibinfo {author} {\bibfnamefont {L.}~\bibnamefont {Brey}},\
  }\href {\doibase 10.1103/PhysRevLett.97.116805} {\bibfield  {journal}
  {\bibinfo  {journal} {Phys. Rev. Lett.}\ }\textbf {\bibinfo {volume} {97}},\
  \bibinfo {pages} {116805} (\bibinfo {year} {2006})}\BibitemShut {NoStop}%
\bibitem [{\citenamefont {Shimshoni}\ \emph {et~al.}(2009)\citenamefont
  {Shimshoni}, \citenamefont {Fertig},\ and\ \citenamefont {Pai}}]{Fertig2009}%
  \BibitemOpen
  \bibfield  {author} {\bibinfo {author} {\bibfnamefont {E.}~\bibnamefont
  {Shimshoni}}, \bibinfo {author} {\bibfnamefont {H.~A.}\ \bibnamefont
  {Fertig}}, \ and\ \bibinfo {author} {\bibfnamefont {G.~V.}\ \bibnamefont
  {Pai}},\ }\href {\doibase 10.1103/PhysRevLett.102.206408} {\bibfield
  {journal} {\bibinfo  {journal} {Phys. Rev. Lett.}\ }\textbf {\bibinfo
  {volume} {102}},\ \bibinfo {pages} {206408} (\bibinfo {year}
  {2009})}\BibitemShut {NoStop}%
\bibitem [{\citenamefont {Murthy}\ \emph {et~al.}(2014)\citenamefont {Murthy},
  \citenamefont {Shimshoni},\ and\ \citenamefont {Fertig}}]{Fertig2014}%
  \BibitemOpen
  \bibfield  {author} {\bibinfo {author} {\bibfnamefont {G.}~\bibnamefont
  {Murthy}}, \bibinfo {author} {\bibfnamefont {E.}~\bibnamefont {Shimshoni}}, \
  and\ \bibinfo {author} {\bibfnamefont {H.~A.}\ \bibnamefont {Fertig}},\
  }\href {\doibase 10.1103/PhysRevB.90.241410} {\bibfield  {journal} {\bibinfo
  {journal} {Phys. Rev. B}\ }\textbf {\bibinfo {volume} {90}},\ \bibinfo
  {pages} {241410} (\bibinfo {year} {2014})}\BibitemShut {NoStop}%
\bibitem [{\citenamefont {Murthy}\ \emph {et~al.}(2016)\citenamefont {Murthy},
  \citenamefont {Shimshoni},\ and\ \citenamefont {Fertig}}]{Fertig2016a}%
  \BibitemOpen
  \bibfield  {author} {\bibinfo {author} {\bibfnamefont {G.}~\bibnamefont
  {Murthy}}, \bibinfo {author} {\bibfnamefont {E.}~\bibnamefont {Shimshoni}}, \
  and\ \bibinfo {author} {\bibfnamefont {H.~A.}\ \bibnamefont {Fertig}},\
  }\href {\doibase 10.1103/PhysRevB.93.045105} {\bibfield  {journal} {\bibinfo
  {journal} {Phys. Rev. B}\ }\textbf {\bibinfo {volume} {93}},\ \bibinfo
  {pages} {045105} (\bibinfo {year} {2016})}\BibitemShut {NoStop}%
\bibitem [{\citenamefont {Tikhonov}\ \emph {et~al.}(2016)\citenamefont
  {Tikhonov}, \citenamefont {Shimshoni}, \citenamefont {Fertig},\ and\
  \citenamefont {Murthy}}]{Fertig2016b}%
  \BibitemOpen
  \bibfield  {author} {\bibinfo {author} {\bibfnamefont {P.}~\bibnamefont
  {Tikhonov}}, \bibinfo {author} {\bibfnamefont {E.}~\bibnamefont {Shimshoni}},
  \bibinfo {author} {\bibfnamefont {H.~A.}\ \bibnamefont {Fertig}}, \ and\
  \bibinfo {author} {\bibfnamefont {G.}~\bibnamefont {Murthy}},\ }\href
  {\doibase 10.1103/PhysRevB.93.115137} {\bibfield  {journal} {\bibinfo
  {journal} {Phys. Rev. B}\ }\textbf {\bibinfo {volume} {93}},\ \bibinfo
  {pages} {115137} (\bibinfo {year} {2016})}\BibitemShut {NoStop}%
\bibitem [{\citenamefont {Young}\ \emph {et~al.}(2014)\citenamefont {Young},
  \citenamefont {Sanchez-Yamagishi}, \citenamefont {Hunt}, \citenamefont
  {Choi}, \citenamefont {Watanabe}, \citenamefont {Taniguchi}, \citenamefont
  {Ashoori},\ and\ \citenamefont {Jarillo-Herrero}}]{Young2014}%
  \BibitemOpen
  \bibfield  {author} {\bibinfo {author} {\bibfnamefont {A.~F.}\ \bibnamefont
  {Young}}, \bibinfo {author} {\bibfnamefont {J.~D.}\ \bibnamefont
  {Sanchez-Yamagishi}}, \bibinfo {author} {\bibfnamefont {B.}~\bibnamefont
  {Hunt}}, \bibinfo {author} {\bibfnamefont {S.~H.}\ \bibnamefont {Choi}},
  \bibinfo {author} {\bibfnamefont {K.}~\bibnamefont {Watanabe}}, \bibinfo
  {author} {\bibfnamefont {T.}~\bibnamefont {Taniguchi}}, \bibinfo {author}
  {\bibfnamefont {R.~C.}\ \bibnamefont {Ashoori}}, \ and\ \bibinfo {author}
  {\bibfnamefont {P.}~\bibnamefont {Jarillo-Herrero}},\ }\href@noop {}
  {\bibfield  {journal} {\bibinfo  {journal} {Nature}\ }\textbf {\bibinfo
  {volume} {505}},\ \bibinfo {pages} {528} (\bibinfo {year}
  {2014})}\BibitemShut {NoStop}%
\bibitem [{\citenamefont {Herbut}(2007)}]{Herbut2007}%
  \BibitemOpen
  \bibfield  {author} {\bibinfo {author} {\bibfnamefont {I.~F.}\ \bibnamefont
  {Herbut}},\ }\href {\doibase 10.1103/PhysRevB.75.165411} {\bibfield
  {journal} {\bibinfo  {journal} {Phys. Rev. B}\ }\textbf {\bibinfo {volume}
  {75}},\ \bibinfo {pages} {165411} (\bibinfo {year} {2007})}\BibitemShut
  {NoStop}%
\bibitem [{\citenamefont {Gusynin}\ \emph {et~al.}(2006)\citenamefont
  {Gusynin}, \citenamefont {Miransky}, \citenamefont {Sharapov},\ and\
  \citenamefont {Shovkovy}}]{Miransky2006}%
  \BibitemOpen
  \bibfield  {author} {\bibinfo {author} {\bibfnamefont {V.~P.}\ \bibnamefont
  {Gusynin}}, \bibinfo {author} {\bibfnamefont {V.~A.}\ \bibnamefont
  {Miransky}}, \bibinfo {author} {\bibfnamefont {S.~G.}\ \bibnamefont
  {Sharapov}}, \ and\ \bibinfo {author} {\bibfnamefont {I.~A.}\ \bibnamefont
  {Shovkovy}},\ }\href {\doibase 10.1103/PhysRevB.74.195429} {\bibfield
  {journal} {\bibinfo  {journal} {Phys. Rev. B}\ }\textbf {\bibinfo {volume}
  {74}},\ \bibinfo {pages} {195429} (\bibinfo {year} {2006})}\BibitemShut
  {NoStop}%
\bibitem [{\citenamefont {Gusynin}\ \emph {et~al.}(2008)\citenamefont
  {Gusynin}, \citenamefont {Miransky}, \citenamefont {Sharapov},\ and\
  \citenamefont {Shovkovy}}]{Miransky2008}%
  \BibitemOpen
  \bibfield  {author} {\bibinfo {author} {\bibfnamefont {V.~P.}\ \bibnamefont
  {Gusynin}}, \bibinfo {author} {\bibfnamefont {V.~A.}\ \bibnamefont
  {Miransky}}, \bibinfo {author} {\bibfnamefont {S.~G.}\ \bibnamefont
  {Sharapov}}, \ and\ \bibinfo {author} {\bibfnamefont {I.~A.}\ \bibnamefont
  {Shovkovy}},\ }\href {\doibase 10.1103/PhysRevB.77.205409} {\bibfield
  {journal} {\bibinfo  {journal} {Phys. Rev. B}\ }\textbf {\bibinfo {volume}
  {77}},\ \bibinfo {pages} {205409} (\bibinfo {year} {2008})}\BibitemShut
  {NoStop}%
\bibitem [{\citenamefont {Roy}\ \emph {et~al.}(2014)\citenamefont {Roy},
  \citenamefont {Kennett},\ and\ \citenamefont {Das~Sarma}}]{DasSarma2014}%
  \BibitemOpen
  \bibfield  {author} {\bibinfo {author} {\bibfnamefont {B.}~\bibnamefont
  {Roy}}, \bibinfo {author} {\bibfnamefont {M.~P.}\ \bibnamefont {Kennett}}, \
  and\ \bibinfo {author} {\bibfnamefont {S.}~\bibnamefont {Das~Sarma}},\ }\href
  {\doibase 10.1103/PhysRevB.90.201409} {\bibfield  {journal} {\bibinfo
  {journal} {Phys. Rev. B}\ }\textbf {\bibinfo {volume} {90}},\ \bibinfo
  {pages} {201409} (\bibinfo {year} {2014})}\BibitemShut {NoStop}%
\bibitem [{\citenamefont {Lado}\ and\ \citenamefont
  {Fern\'andez-Rossier}(2014)}]{FernandoRossier2014}%
  \BibitemOpen
  \bibfield  {author} {\bibinfo {author} {\bibfnamefont {J.~L.}\ \bibnamefont
  {Lado}}\ and\ \bibinfo {author} {\bibfnamefont {J.}~\bibnamefont
  {Fern\'andez-Rossier}},\ }\href {\doibase 10.1103/PhysRevB.90.165429}
  {\bibfield  {journal} {\bibinfo  {journal} {Phys. Rev. B}\ }\textbf {\bibinfo
  {volume} {90}},\ \bibinfo {pages} {165429} (\bibinfo {year}
  {2014})}\BibitemShut {NoStop}%
\bibitem [{\citenamefont {Lukose}\ and\ \citenamefont
  {Shankar}(2016)}]{Shankar2016}%
  \BibitemOpen
  \bibfield  {author} {\bibinfo {author} {\bibfnamefont {V.}~\bibnamefont
  {Lukose}}\ and\ \bibinfo {author} {\bibfnamefont {R.}~\bibnamefont
  {Shankar}},\ }\href {\doibase 10.1103/PhysRevB.94.085135} {\bibfield
  {journal} {\bibinfo  {journal} {Phys. Rev. B}\ }\textbf {\bibinfo {volume}
  {94}},\ \bibinfo {pages} {085135} (\bibinfo {year} {2016})}\BibitemShut
  {NoStop}%
\bibitem [{\citenamefont {Wehling}\ \emph {et~al.}(2011)\citenamefont
  {Wehling}, \citenamefont {\ifmmode \mbox{\c{S}}\else \c{S}\fi{}a\ifmmode
  \mbox{\c{s}}\else \c{s}\fi{}\ifmmode \imath \else \i
  \fi{}o\ifmmode~\breve{g}\else \u{g}\fi{}lu}, \citenamefont {Friedrich},
  \citenamefont {Lichtenstein}, \citenamefont {Katsnelson},\ and\ \citenamefont
  {Bl\"ugel}}]{Katsnelson2011}%
  \BibitemOpen
  \bibfield  {author} {\bibinfo {author} {\bibfnamefont {T.~O.}\ \bibnamefont
  {Wehling}}, \bibinfo {author} {\bibfnamefont {E.}~\bibnamefont {\ifmmode
  \mbox{\c{S}}\else \c{S}\fi{}a\ifmmode \mbox{\c{s}}\else \c{s}\fi{}\ifmmode
  \imath \else \i \fi{}o\ifmmode~\breve{g}\else \u{g}\fi{}lu}}, \bibinfo
  {author} {\bibfnamefont {C.}~\bibnamefont {Friedrich}}, \bibinfo {author}
  {\bibfnamefont {A.~I.}\ \bibnamefont {Lichtenstein}}, \bibinfo {author}
  {\bibfnamefont {M.~I.}\ \bibnamefont {Katsnelson}}, \ and\ \bibinfo {author}
  {\bibfnamefont {S.}~\bibnamefont {Bl\"ugel}},\ }\href {\doibase
  10.1103/PhysRevLett.106.236805} {\bibfield  {journal} {\bibinfo  {journal}
  {Phys. Rev. Lett.}\ }\textbf {\bibinfo {volume} {106}},\ \bibinfo {pages}
  {236805} (\bibinfo {year} {2011})}\BibitemShut {NoStop}%
\bibitem [{\citenamefont {C\^ot\'e}\ and\ \citenamefont
  {MacDonald}(1991)}]{MacDonald1991}%
  \BibitemOpen
  \bibfield  {author} {\bibinfo {author} {\bibfnamefont {R.}~\bibnamefont
  {C\^ot\'e}}\ and\ \bibinfo {author} {\bibfnamefont {A.~H.}\ \bibnamefont
  {MacDonald}},\ }\href {\doibase 10.1103/PhysRevB.44.8759} {\bibfield
  {journal} {\bibinfo  {journal} {Phys. Rev. B}\ }\textbf {\bibinfo {volume}
  {44}},\ \bibinfo {pages} {8759} (\bibinfo {year} {1991})}\BibitemShut
  {NoStop}%
\bibitem [{\citenamefont {Wang}\ \emph {et~al.}(2008)\citenamefont {Wang},
  \citenamefont {Iyengar}, \citenamefont {Fertig},\ and\ \citenamefont
  {Brey}}]{Jianhui2008}%
  \BibitemOpen
  \bibfield  {author} {\bibinfo {author} {\bibfnamefont {J.}~\bibnamefont
  {Wang}}, \bibinfo {author} {\bibfnamefont {A.}~\bibnamefont {Iyengar}},
  \bibinfo {author} {\bibfnamefont {H.~A.}\ \bibnamefont {Fertig}}, \ and\
  \bibinfo {author} {\bibfnamefont {L.}~\bibnamefont {Brey}},\ }\href {\doibase
  10.1103/PhysRevB.78.165416} {\bibfield  {journal} {\bibinfo  {journal} {Phys.
  Rev. B}\ }\textbf {\bibinfo {volume} {78}},\ \bibinfo {pages} {165416}
  (\bibinfo {year} {2008})}\BibitemShut {NoStop}%
\end{thebibliography}
\end{document}